\documentclass{iopart}
\usepackage{iopams,leftidx,setstack,graphicx,color}
\begin{document}

\title[Transport properties of quantum dots in the Wigner molecule
regime]{Transport properties of quantum dots in the Wigner molecule
  regime}

\author{F Cavaliere$^{1}$, U De Giovannini$^{2}$, M Sassetti$^{1}$,
  and B Kramer$^{2,3}$}
\address{$^{1}$ CNR-INFM LAMIA, Dipartimento di Fisica, Universit\`a di Genova, Via Dodecaneso 33, 16146 Genova Italy\\
  $^{2}$ School of Engineering and Sciences, Jacobs University Bremen,
  Campus Ring 1, 28759 Bremen Germany\\ $^{3}$ I Institut f\"ur
  Theoretische Physik, Universit\"at Hamburg, Jungiusstra\ss e 9, 20355
  Hamburg Germany}
\ead{cavalier@fisica.unige.it}

\begin{abstract}
  The transport properties of quantum dots with up to $N=7$ electrons
  ranging from the weak to the strong interacting regime are
  investigated via the projected Hartree-Fock technique. As
  interactions increase radial order develops in the dot, with the
  formation of ring and centered-ring structures. Subsequently,
  angular correlations appear, signalling the formation of a Wigner
  molecule state.\\
\noindent We show striking signatures of the emergence of Wigner
  molecules, detected in transport. In the linear regime, conductance
  is exponentially suppressed as the interaction strength grows. A
  further suppression is observed when centered-ring structures
  develop, or peculiar spin textures appear. In the nonlinear regime,
  the formation of molecular states may even lead to a conductance
  enhancement.
\end{abstract}

\submitto{\NJP}
\maketitle

\section{Introduction}
Semiconductor quantum dots (QDs), frequently referred to as artificial
atoms, are nanometer-sized structures whose conduction electrons are
confined in all the three spatial
dimensions~\cite{kouw2,Reimann,Landman1}. In these systems a
two-dimensional electron gas, formed at the interface of a
heterojunction, is depleted by chemical etching or electrostatic
potentials in order to form an isolated region, connected to external
reservoirs by tunnel barriers. For a small number of particles $N$,
the potential can often be considered as
harmonic~\cite{kouw2,Reimann,Landman1}.\\
\noindent In analogy to atomic systems, quantum dots can be probed
optically by studying their absorption or emission
spectrum~\cite{Delerue}. Additionally, the study of transport
properties is a source of information for quantum dots embedded into
an electronic circuit~\cite{kouw2}. The current flow proceeds by
tunnelling events once a bias voltage is applied to the external
reservoirs and the presence of an external gate voltage allows to tune
the number of excess electrons in the dot with respect to a neutral
configuration.\\

\noindent Theoretically, the study of correlated quantum dot states is
a challenging many-body problem: even the fairly simple case of $N=2$
can be solved exactly only in specific regimes~\cite{Taut}, whereas in
general one has to resort to semi-analytic methods~\cite{2D1} or
approximate WFs~\cite{2D2}. For $N>2$, several numerical methods have
been employed. In increasing order of computational complexity they
range from unrestricted Hartree-Fock
methods~\cite{yann1,yann6,reush1,szafran0} and density functional
theory~\cite{gattobigio1,kosinen,esa1,esa2,esa3}, to {\em projected}
Hartree-Fock
(PHF)~\cite{koonin,yann3,yann6,yann4,yann7,PHF2,PHF,PHF3}, random
phase approximation~\cite{RPA1,RPA2}, quantum Monte
Carlo~\cite{esa3,egger1,ghosal1,ghosal2,ghosal3}, and exact
diagonalization~\cite{Pfan1,Mak1,yann2,mik1,mik2,mik3,Tav1,Tav2,Tav3,RontaniCI}. Recently,
PHF techniques have been used by our group for the study of correlated
quantum dots. While retaining the flexibility of an unrestricted
Hartree-Fock approach, PHF allows to overcome the limitations due to
symmetry broken solutions and to efficiently obtain dot wavefunctions
(WFs) with the correct spin and angular momentum, showing correlations
beyond the mean field level~\cite{PHF2,PHF}.

\noindent Circular two-dimensional systems realized e.g. by pillar
quantum dots display interesting features depending on the ratio
$\lambda$ between the strength of the typical interaction and the
strength of the confining potential -- see Sec.~\ref{sec:dot} for a
precise definition of $\lambda$. When $\lambda$ is small, atomic-like
effects due to quantum mechanical confinement such as the formation of
shell structures have been observed~\cite{kouw1,tarucha,sasaki} and
explained at the mean field level~\cite{kouw2,kouw3}. For increasing
$\lambda$, several numerical investigations have shown the emergence
of correlated electron states and the occurrence of Wigner molecular
states~\cite{bedanov,mik1,mik2,Tav1,Tav2,Tav3,RontaniCI,maksym1,egger1,yann1,yann2,yann3,reimann2,reush1,filinov,kosinen1,jean,harju1,mik3,reush2,yann4,szafran1,weiss1,ghosal1,ghosal2,ghosal3,ludwig1,rontani1,maximaging,kalliakos1,zeng1,gattobigio1},
the finite-size analogue of Wigner crystals~\cite{wigner1,vignale},
characterized by correlations {\em beyond} the mean field.\\
\noindent The transition towards a molecular state occurs smoothly~\cite{egger1,gattobigio1,ghosal1,ghosal2,ghosal3,filinov,zeng1}. As
interactions increase, the dot WFs cross over from weakly correlated
states at small $\lambda$ to Wigner molecular states, characterized by
strong correlations. The transition occurs in two phases:
\begin{itemize}
\item At small $\lambda$, correlations begin to develop and ring-like
  structures develop in the dot WF. For $N\geq 6$, also centered
  structures with one or more electrons in the dot center may form.
\item For higher $\lambda$, angular correlations begin to appear as
  the dot enters the {\em incipient Wigner molecule}
  regime~\cite{ghosal1,ghosal2}. Increasing $\lambda$ further, the dot
  WF represents a rotating Wigner molecule~\cite{maksym1} and the
  electrons localize around the equilibrium positions of a classical
  Coulomb molecule~\cite{bedanov,jean}.
\end{itemize}
\noindent Correlations are particularly relevant also in
one-dimensional systems which display an analogous transition towards
the Wigner
molecule~\cite{Jauregui,szafran99,polini1,polini2,sechi1}.\\
\noindent The experimental observation of strongly correlated states
in quantum dots has attracted considerable interest. In pillar quantum
dots, inelastic light scattering experiments have shown signatures of
correlated quantum states~\cite{kalliakos1,garcia1}. Scanning
tunnelling spectroscopy experiments for the imaging of correlated
quantum dot WFs were recently performed and theoretically
analyzed~\cite{rontani1,maximaging,bester}. Also transport properties
can yield information about correlated states. In one dimension, the
influence of correlations on the transport properties is predicted to
be particularly important~\cite{kleimann1,cava1,cava2}. Recently,
experimental evidence of the formation of few-electrons Wigner
molecules has been reported in carbon nanotubes~\cite{desphande1}.

\noindent Also spin correlations can heavily influence the transport
properties of quantum dots, even in the absence of an applied magnetic
field. In quantum dots with asymmetric tunnel barriers, the degeneracy
of spin multiplets may lead to asymmetric current-voltage
characteristics~\cite{akera}. Another notable example is the type-II
spin blockade~\cite{wein1,wein2}, which occurs in the linear transport
regime when the absolute value of the difference between the total
spin of initial and final dot states
exceeds 1/2 and leads to zero sequential current through the dot.\\

\noindent In this paper we investigate the transport properties of
quantum dots in the presence of strong correlations. In such a regime,
a mean field treatment in the spirit of the so called ``constant
interaction model''~\cite{kouw1,beenakker} is clearly not
viable. Indeed, one has to resort to more precise techniques to obtain
the spectrum and the WFs. Numerical studies of the transport
properties, similar to the one proposed here have been performed in
the past employing exact diagonalizations for $N\leq3$ electrons in
circular QDs~\cite{pfannkuche,uzi1,melnikov} and $N\leq4$ electrons in
one-dimensional quantum dots~\cite{jauregui1}. These works, however,
were not focused on the signatures due to Wigner molecules in the
transport properties.\\

\noindent In the present work we numerically investigate the transport
properties of pillar quantum dots beyond the constant interaction
model. Our model is that of $N$ interacting electrons confined to a
two-dimensional plane and further subject to an in-plane harmonic
potential. More refined models, including effects due to a finite
thickness of the dot and to heavy doping in the reservoirs, have been
recently proposed~\cite{Maksym}. In this work we will neglect such
effects, addressing systems in which the screening is moderate (a
strong screening may hinder the formation of Wigner
molecules~\cite{Jauregui,screen1}). We use the PHF method in order to
estimate the correlated dot WFs for $4\leq N\leq7$ in a range of
$\lambda$ which allows to observe the transition between liquid-like
and molecular electron states. Sequential tunnelling rates are
numerically evaluated and the dot conductance is obtained using a rate
equation.\\

\noindent Our task is to understand whether or not peculiar signatures in the
transport properties may be detected as a consequence of the
transition towards the Wigner molecule. According to the results
presented in this paper, the answer is affirmative.\\

\noindent In the linear transport regime, qualitative modifications of
the dot ground state (GS) WFs induce a peculiar suppression of the
conductance. Such qualitative modifications may be induced either by
the formation of centered ring-like structures or by the emergence of
peculiar spin patterns in the dot WF. Both cases are presented in this
paper.\\
\noindent Signatures of the transition can also be seen in the
nonlinear transport regime. We have found that the tunnelling rate
through an excited state of the dot may be increased strongly by the
formation of a Wigner molecule.\\
\noindent The features described above are genuine hallmarks of the
formation of Wigner molecules in the dot and can be expected to be
observable in experiments.\\

\noindent The outline of the paper is as follows. In
Sec.~\ref{sec:modelandmethods} we introduce the model and the PHF
method, we discuss the tunnelling Hamiltonian and the rate equation
for calculating the current. Results are presented in
Sec.~\ref{sec:results}. Here, after discussing in detail the
occurrence of Wigner molecules, we show results for the conductance in
both the linear and nonlinear regimes. Conclusions are presented in
Sec.~\ref{sec:conclusions}. \ref{sec:a1} contains the derivation of
the tunnelling Hamiltonian while the dot tunnelling rates are
discussed in~\ref{sec:a2}.
\section{Model and methods}
\label{sec:modelandmethods}
\subsection{Quantum dot}
\label{sec:dot}
\begin{figure}[ht]
\begin{center}
\includegraphics[height=8cm,keepaspectratio]{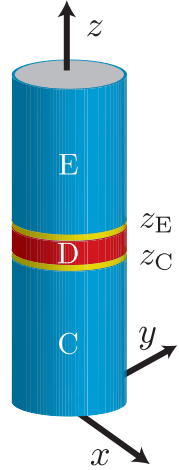}
\vspace{-1.8cm}
\caption{Schematic representation of a pillar quantum dot. The red
  thin disc is the quantum dot (D), connected via tunneling barriers
  at $z\approx z_{\mathrm E},z_{\mathrm C}$ (yellow parts) to emitter
  (E) and collector (C) leads, represented in blue.}
\label{fig:fig1}
\end{center}
\end{figure}

In a pillar quantum dot~\cite{kouw1,tarucha,sasaki,kouw3} electrons
are confined to a thin disk of semiconducting material, represented by
the red region in figure~\ref{fig:fig1}. The dot is embedded between
tunnel barriers located around $z=z_{\mathrm E},z_{\mathrm C}$, with
$z$ the axial direction. The tunnel barriers couple the dot to the
external emitter and collector contacts~\cite{Reimann}. A metallic
gate is assumed to surround the dot region (not shown in
figure~\ref{fig:fig1}) and allows to shift the dot energy levels as a
suitable gate voltage $V_{\mathrm g}$ is applied to it.
Due to the strong confinement along $z$, the motion of electrons is
essentially restricted to the $(x,y)$ plane. Electrons are further
subject to a lateral confining potential with rotational symmetry
around the $z$ axis, as appropriate for the cylindrical quantum dots
studied in this paper. For small dots, containing few electrons, this
potential is well approximated by a parabolic one~\cite{Reimann}. The
Hamiltonian for $N$ interacting electrons is $\hat{H}_{\mathrm
  D}=\hat{H}_{\mathrm D}^{(0)}+\hat{H}_{\mathrm D}^{(1)}$ with
($\hbar=1$, boldface denotes vectors)
\begin{equation}
\label{eq:dot1}
\fl\hat{H}_{\mathrm D}^{(0)}=\sum_{i=1}^{N}\left[\frac{1}{2m^{*}}\hat{{\mathbf P}}_{i}^{2}+\frac{m^{*}\omega^{2}}{2}\hat{{\mathbf R}}_{i}^{2}\right]-eN V_{\mathrm g}\ \ ;\ \ \hat{H}_{\mathrm D}^{(1)}=\frac{e^{2}}{4\pi\varepsilon_{0}\varepsilon_{{\mathrm r}}}\sum_{i=1}^{N-1}\sum_{j=i+1}^{N}\frac{1}{|\hat{\mathbf{R}}_{i}-\hat{\mathbf{R}}_{j}|}\, ,
\end{equation}
where $\hat{\mathbf{R}}_{i}=(\hat{x}_{i},\hat{y}_{i})$ is the $i$-th
electron coordinate and $\hat{\mathbf{P}}_{i}$ its momentum. Here,
$-e$ and $m^{*}$ are the electron charge and effective mass
respectively. Furthermore, $\omega$ is the confinement energy,
$\varepsilon_{0}$ ($\varepsilon_{\rm r}$) the vacuum (relative)
dielectric constant. We will consider a bare Coulomb potential for the
interaction term $\hat{H}_{\mathrm D}^{(1)}$, neglecting both
finite-thickness effects and screening due to heavily doped
contacts. Such effects modify the interaction potential producing
deviations from the $r^{-1}$ behaviour for both short and long
inter-electron distances $r$~\cite{Maksym}. Finite thickness effects
would also produce a renormalization of the gate voltage $V_{\mathrm
  g}$~\cite{Maksym}. Our calculations are therefore valid
for systems characterized by weak screening.\\
\noindent Expressing lengths in units of
$\ell_{0}=(m^{*}\omega)^{-1/2}$ and energies in units $\omega$, the
Hamiltonian becomes
\begin{equation}
  \hat{H}_{\mathrm D}^{(0)}=\sum_{i=1}^{N}\left[\frac{\hat{{\mathbf
          p}}_{i}^{2}}{2}+\frac{\hat{{\mathbf
          r}}_{i}^{2}}{2}\right]-\frac{eV_{\mathrm
      g}}{\omega_{0}}N\quad;\quad\hat{H}_{\mathrm
    D}^{(1)}=\lambda\sum_{i=1}^{N-1}\sum_{j=i+1}^{N}\frac{1}{|\hat{\mathbf{r}}_{i}-\hat{\mathbf{r}}_{j}|}\label{eq:dotB0}\,
    ,
\end{equation}
with $\hat{\mathbf{r}}_{i}=\hat{\mathbf{R}}_{i}/\ell_{0}$ and
$\hat{\mathbf{p}}_{i}=\hat{\mathbf{P}}_{i}\ell_{0}$. The dimensionless
parameter
\begin{equation}
  \lambda=\frac{e^{2}}{4\pi\varepsilon_{0}\varepsilon_{\rm r}\ell_{0}\omega}=\frac{\ell_{0}}{a_{\mathrm B}^{*}}\label{eq:lambda}
\end{equation}
measures the Coulomb interaction strength. It is the ratio between the
effective length scale $\ell_{0}$ and the effective Bohr radius
$a_{\mathrm B}^{*}=4\pi\varepsilon_{0}\varepsilon_{\rm
  r}/m^{*}e^{2}$. Experimentally, the interaction strength $\lambda$
can be modified by tuning the confinement strength $\omega$ via
electrostatic gates.\\
\noindent In the rest of the paper, we will concentrate on GaAs
quantum dots, where $\varepsilon_{\mathrm r}=12.4$ and $m^{*}=0.067
m_{\mathrm e}$ with $m_{\mathrm e}=9.1\cdot 10^{-31}$ kg. In this
case, expressing $\omega$ in meV, one has $\lambda\approx
3.46\ \sqrt{\mathrm{meV}}/\sqrt{\omega}$. Weak (strong) interactions occur
for $\lambda\lesssim 1$ ($\lambda>1$). In the absence of interactions
($\lambda=0$) the problem can be solved exactly~\cite{FD}. The
eigenstates of $\hat{H}_{\mathrm D}^{(0)}$ are Fock-Darwin (FD) states
labelled by a principal quantum number $n\geq0$, by the electron
angular momentum ($z$ component) $l\in\mathbb{Z}$ and by the electron
spin $z$ component $s_{z}=\pm1/2$. The corresponding WFs are denoted
by $f_{n,l,s_{z}}(\mathbf{r})$ and the spin degenerate energy spectrum
is given by $E_{n,l,s_{z}}=\omega(2n+|l|+1)$. In the presence of
interactions, the problem cannot be tackled analytically if $N>2$ and
one has to use numerical techniques. It is important to notice the
symmetries of $\hat{H}_{\mathrm D}$: it commutes with the total
angular momentum ($z$ component) $\hat{L}$, the total spin
$\hat{\mathbf{S}}$ and the total spin $z$ component $\hat{S}_{z}$. As
a consequence, their eigenvalues can be used to label the dot energy
spectrum and WFs. These are obtained by means of the PHF technique
which has been extensively described in~\cite{PHF2,PHF}. Here, we
briefly outline the procedure. For a given particle number $N$ and
each value of $-N/2\leq S_{z}\leq N/2$, the dot WFs are first
approximated as single Slater determinants $|N,S_{z}\rangle$ made up
of $N_{\uparrow}$ ($N_{\downarrow}$) orbital with spin $s_{z}=1/2$
($s_{z}=-1/2$) where $S_{z}=(N_{\uparrow}-N_{\downarrow})/2$ and
$N=N_{\uparrow}+N_{\downarrow}$. Orbitals are variationally optimized
with the spin and spatially unrestricted Hartree-Fock
method~\cite{yann1} which produces several stationary states
$|N,S_{z}\rangle_{i}$, in general neither eigenstates of $\hat{L}$,
nor of $\hat{S}^{2}$. Projection operators $\hat{P}_{L,S}$ are
subsequently applied to $|N,S_{z}\rangle_{i}$ to restore the
symmetries broken due to the single Slater determinant ansatz. As a
result, correlated WFs
\begin{equation}
\label{eq:phfwf}
|N,L,S,S_{z}\rangle_{i}=\hat{P}_{L,S}|N,S_{z}\rangle_{i}\nonumber
\end{equation}
are obtained. The state in~(\ref{eq:phfwf}) cannot be represented
as a single Slater determinant and contains correlations beyond mean
field. The dot ground state is obtained as the state which minimizes
the energy
\begin{equation}
E_{N,L_{0},S_{0},S_{z0}}=\min_{L,S,S_{z},i}\left\{\frac{\leftidx{_i}{\langle N,S_{z}|}\hat{H}\hat{P}_{L,S}|N,S_{z}\rangle_{i}}
{\leftidx{_i}{\langle N,S_{z}|}\hat{P}_{L,S}|N,S_{z}\rangle_{i}}
\right\}\, .\label{eq:gs}
\end{equation}
Here, $L_{0}$, $S_{0}$, $S_{z0}$ is the set of quantum numbers which
minimize~(\ref{eq:gs}) and which label the dot ground state for $N$
electrons, with WF $|N,L_{0},S_{0},S_{z0}\rangle$.  In a
similar fashion one defines excited states within the PHF method. For
instance, the first excited state is given by
\begin{equation}
E_{N,L_{1},S_{1},S_{z1}}=\min_{L,S,S_{z},i}\left\{\frac{\leftidx{_i}{\langle N,S_{z}|}\hat{H}\hat{P}_{L,S}|N,S_{z}\rangle_{i}}
{\leftidx{_i}{\langle N,S_{z}|}\hat{P}_{L,S}|N,S_{z}\rangle_{i}}>E_{N,L_{0},S_{0},S_{z0}}
\right\}\, ,\label{eq:1ex}
\end{equation}
with $L_{1}$, $S_{1}$, and $S_{z1}$ determined by the minimization
procedure. As a consequence of the correlations introduced by the
projection technique, energies {\em lower} than those obtained by
unrestricted Hartree-Fock are achieved~\cite{PHF2,PHF}.
\subsection{Tunnelling rates}
As shown in~\ref{sec:a1}, assuming cylindrical symmetry about the $z$
axis and separability of longitudinal and transverse motions, the
tunnelling Hamiltonian between the quantum dot and the leads
is~\cite{Bardeen}
\begin{equation}
\label{eq:htun}
\hat{H}_{\mathrm t}=\sum_{\alpha=\mathrm{E,C}}\sum_{\xi_{\alpha},\eta}\tau_{\xi_{\alpha},\eta}^{(\alpha)}\hat{c}_{\alpha,\xi_{\alpha}}^{\dagger}\hat{d}_{\eta}+\mathrm{h.c.}
\end{equation}
where $\alpha={\mathrm E}$ ($\alpha={\mathrm C}$) for the emitter
(collector), $\xi_{\alpha}$ and $\eta$ collectively denote a set of
single particle quantum numbers for the lead $\alpha$ and the dot
respectively and $\tau_{\xi_{\alpha},\eta}^{(\alpha)}$ is the
tunnelling amplitude. In the following, we will choose the FD states
$\eta=\left\{n_{\eta}^{(\mathrm{D})},l_{\eta}^{(\mathrm{D})},s_{z\eta}^{(\mathrm{D})}\right\}$
as a basis of single particle states for the dot. In the case of a
pillar quantum dot one obtains $\tau_{\xi_{\alpha},\eta}\approx
t^{(\alpha)}\delta_{\nu_{\alpha},\eta} $, see~(\ref{eq:tau}). The
choice of FD states is not restrictive: indeed every orthonormal and
complete basis for the single particle states of the dot produces
identical results, as shown in~\ref{sec:a3}. The fermionic operator
for the lead $\alpha$ is $\hat{c}_{\alpha,\xi_{\alpha}}$, while
$\hat{d}_{\eta}$ is the one for the dot. Leads are treated as
noninteracting Fermi gases with the Hamiltonian
\begin{equation}
\label{eq:hleads}
\hat{H}_{\alpha}=\sum_{\xi_{\alpha}}E_{\alpha}(\xi_{\alpha})c_{\alpha,\xi_{\alpha}}^{\dagger}c_{\alpha,\xi_{\alpha}}\, ,
\end{equation}
with energy spectrum $E_{\alpha}(\xi_{\alpha})$.\\
\noindent Our task is to evaluate the sequential tunnelling rates
between initial ($|\mathcal{I}_{\mathrm D}\rangle$) and final
($|\mathcal{F}_{\mathrm D}\rangle$) dot states with energies
$E_{{\mathcal I}_{\mathrm D}}$ and $E_{{\mathcal F}_{\mathrm D}}$,
respectively. As shown in~\ref{sec:a2}, the rates are obtained by
tracing out the degrees of freedom of the leads, and have the general
form
\begin{equation}
\label{eq:tunnelingrate}
\Gamma_{\mathcal{I}_{\mathrm D}\to\mathcal{F}_{\mathrm D}}=\sum_{\alpha=\mathrm{E,C}}\sum_{p=\pm 1}\Gamma_{\mathcal{I}_{\mathrm D}\to\mathcal{F}_{\mathrm D}}^{(\alpha),p}\, ,
\end{equation}
where $p=+1$ ($p=-1$) represents tunnelling into (out from) the dot via
lead $\alpha$. They are
\begin{equation}
\label{eq:ratepil} 
\Gamma_{\mathcal{I}_{\mathrm D}\to\mathcal{F}_{\mathrm
D}}^{(\alpha),p}=\Gamma^{(\alpha)}\left|\mathcal{O}_{p}\right|^{2}f_{p}(
\mu_{\mathrm D}-\mu_{\alpha})\, ,
\end{equation} 
where $\Gamma^{(\alpha)}=2\pi\mathcal{D}_{\alpha}|t^{(\alpha)}|^{2}$
is the bare tunnelling rate with $\mathcal{D}_{\alpha}$ the density of
states of lead $\alpha$ and $f_{p}(E)=p f(E)+(1-p)/2$, where
$f(E)=[1+\exp(\beta E)]^{-1}$ is the Fermi distribution at inverse
temperature $\beta=1/k_{\mathrm B}T$ ($k_{\mathrm B}$ the Boltzmann
constant). The chemical potential of the dot is $\mu_{\mathrm
  D}=E_{{\mathcal F}_{\mathrm D}}-E_{{\mathcal I}_{\mathrm D}}$ and
those for the leads are
$\mu_{\alpha}=\mu_{0}+\delta\mu_{\alpha}$. Here, $\delta\mu_{\alpha}$
is a shift due to the presence of a bias voltage $V$. In the
following, symmetric voltage drops will be assumed at the barriers,
with $\delta\mu_{\mathrm E}=eV/2$ and $\delta\mu_{\mathrm C}=-eV/2$.\\
\noindent Interaction effects are embodied into the term
\begin{equation}
\label{eq:amplitudes}
\mathcal{O}_{1}=\sum_{\eta}\langle\mathcal{F}_{\mathrm D}|\hat{d}_{\eta}^{\dagger}|\mathcal{I}_{\mathrm D}\rangle\quad;\quad\mathcal{O}_{-1}=\sum_{\eta}\langle\mathcal{F}_{\mathrm D}|\hat{d}_{\eta}|\mathcal{I}_{\mathrm D}\rangle\label{eq:matelinout}\, , 
\end{equation} 
which can be evaluated numerically once the initial and the final dot
states have been obtained by means of PHF. For $\lambda=0$, one can
only have $|\mathcal{O}_{p}|^2=0,1$ depending on the initial and final
dot states. For $\lambda>0$, on the other hand,
$|\mathcal{O}_{p}|^{2}$ is not limited to these two extreme
cases. Note that $|\mathcal{O}_{p}|^2$ contains interference effects
between different FD orbitals.\\
\subsection{Rate equation}
Using the tunnelling rates one can set up a rate equation for the
occupation probabilities $P_{\mathcal {I}}$ of the dot states
$|\mathcal{I}\rangle$ (in this section, we omit the subscript D for
simplicity)
\begin{equation} 
\label{eq:rateq}
\frac{\partial{P}_{\mathcal{I}}}{\partial t}=\sum_{\mathcal{J}}M_{\mathcal{I}\mathcal{J}}P_{\mathcal{J}}\, .
\end{equation}  
The rate equation is a powerful and standard tool to study the
transport properties of quantum dots, especially in the sequential
regime~\cite{beenakker}.

\noindent The transition matrix $M_{\mathcal{IJ}}$ is defined as
\begin{eqnarray*}
M_{\mathcal{I}\mathcal{J}}=\Gamma_{\mathcal{J}\to\mathcal{I}}\left(1-\delta_{N_{\mathcal
I},N_{\mathcal
J}}\right)+W_{\mathcal{J}\to\mathcal{I}}\delta_{N_{\mathcal
I},N_{\mathcal J}}\quad\quad&&\mathrm{if}\
\mathcal{I}\neq\mathcal{J}\\
M_{\mathcal{I}\mathcal{I}}=-\sum_{\mathcal{I'}\neq\mathcal{I}}M_{\mathcal{I'}\mathcal{I}}\,
, \end{eqnarray*} with $\Gamma_{{\mathcal J}\to{\mathcal I}}$ given
in~(\ref{eq:tunnelingrate}). In order to take into account dissipation
effects on the excited states, we introduced a phenomenological
relaxation rate \begin{eqnarray}
W_{\mathcal{I}\to\mathcal{J}}=W\quad\quad&&\mathrm{if}\
E_{\mathcal{J}}<E_{\mathcal{I}}\, ;\nonumber\\
W_{\mathcal{I}\to\mathcal{J}}=W \rme^{-\beta(E_{\mathcal
J}-E_{\mathcal I})}\quad\quad&&\mathrm{if}\ E_{\mathcal{J}}\geq
E_{\mathcal{I}}\, .\label{eq:relaxation} \end{eqnarray} \noindent In
the stationary regime, the left hand side of~(\ref{eq:rateq}) vanishes
and the rate equation reduces to a standard linear system of equations
for the stationary occupation probabilities of dot states
$\bar{P}_{\mathcal I}$ which can be easily solved by means of singular
value decomposition since $\det(M_{\mathcal{I}\mathcal{J}})=0$. The
solution is uniquely determined by imposing the normalization
condition $\sum_{\mathcal
  I}\bar{P}_{\mathcal I}=1$. Once the dot occupation probabilities are
obtained, the stationary current $I^{(\alpha)}$ through barrier
$\alpha$ can be calculated with the aid of the barrier-resolved
tunnelling rates~(\ref{eq:ratepil}) as
\begin{equation} 
  I^{(\alpha)}=e\sum_{\mathcal I}\sum_{\mathcal{J}\neq I}\sum_{p=\pm 1}p\bar{P}_{\mathcal I}\Gamma_{\mathcal{I}\to\mathcal{J}}^{(\alpha),p}\, . \end{equation} 
In the stationary regime, $I^{(\mathrm E)}=-I^{(\mathrm C)}=I$. The differential conductance is defined as $\mathcal{G}=\partial I/\partial V$.
\subsection{Quasiparticle wavefunction}
\label{sec:qpwf}
Useful information about the dot states can also be extracted from
the {\em quasiparticle WF} (QPWF)~\cite{maximaging}
\begin{equation}
\label{eq:qpwfs}
\varphi(\mathbf{r})=\sum_{s_{z}=\pm 1/2}\langle\mathcal{F}_{\mathrm D}|\hat{\psi}^{\dagger}_{s_z}(\mathbf{r})|\mathcal{I}_{\mathrm D}\rangle
\end{equation}
where 
\begin{equation}
\label{eq:fieldop}
\hat{\psi}_{s_z}^{\dagger}(\mathbf{r})=\sum_{n\geq 0,l}f_{n,l,s_z}^{*}(\mathbf{r})\hat{d}_{n,l,s_z}^{\dagger}
\end{equation}
is the dot field operator, with $f_{n,l,s_{z}}(\mathbf{r})$ the FD
WFs, and the final dot state $|\mathcal{F}_{\mathrm D}\rangle$ has one
extra electron with respect to the initial one $|\mathcal{I}_{\mathrm
  D}\rangle$. The squared modulus $|\varphi(\mathbf{r})|^{2}$ is
proportional to the probability density
of tunnelling into the dot at position $\mathbf{r}$.\\
\noindent The QPWF is the analog of the single particle WF of a
tunnelling electron for the case of an interacting quantum dot: for
$\lambda=0$ it simply reduces to the WF of the FD orbital occupied by
the tunnelling electron.\\
\noindent For a transition from the state $|N,L,S,S_{z}\rangle$ to
$|N+1,L',S',S_{z}'\rangle$, the QPWF has the general form in polar
coordinates ${\mathbf r}\to(r,\theta)$~\cite{maximaging}
\begin{eqnarray}
\varphi(r,\theta)=\rme^{\rmi\theta\Delta L}|\varphi(r)|\, ,
\end{eqnarray}
where $\Delta L=L'-L$. Furthermore, $|\varphi(r)|\propto r^{|\Delta
  L|}$ if $r\to 0$ and $|\varphi(r)|\to 0$ for $r\to\infty$.
\section{Results}
\label{sec:results}
In this section we present results for a GaAs-based quantum dot with
$4\leq N\leq 7$, with parameters $\varepsilon_{\mathrm r}=12.4$ and
$m^{*}=0.067m_{\mathrm e}$ where $m_{\mathrm e}=9.1\cdot10^{-31}$ kg.\\
\noindent For the PHF calculations, we use a truncated basis
consisting of 75 FD states per spin direction. Projection operators
are numerically implemented with a fast Fourier transform over 256
samples. For further details, see~\cite{PHF}. The ground state and
first few excited states are obtained for interaction strengths in the
range $1\leq\lambda\leq 2.8$.
\subsection{Molecular states of electrons} 
\label{sec:molecules}
In table~\ref{tab:tab1} the quantum numbers of the many-body ground states of the
dot for increasing values of $\lambda$ are shown as derived from the
PHF procedure. Dot states consist of multiplets, degenerate on $S_{z}$
and on $L=\pm L_{0}$.
\begin{table}
  \caption{\label{tab:tab1}Quantum numbers of the dot GSs as a function of $N$ in the
    range $1\leq\lambda\leq2.8$.}
\begin{indented}
\item[]\begin{tabular}{@{}llll} 
\br
$N$&L&S&$S_{z}$\\ 
\mr 
4&0&1&0,$\pm1$\\ 
5&$\pm 1$&1/2&$\pm1/2$\\
6&0&0&0\\ 
7&$\pm 2$&1/2&$\pm1/2$\\ 
\br 
\end{tabular} 
\end{indented} 
\end{table} 
Dot quantum numbers are constant throughout the whole range of
interaction strengths $1\leq\lambda\leq2.8$ considered in this
paper. They agree with the results of more refined exact
diagonalizations~\cite{RontaniCI}.\\ 
\noindent As the interaction strength increases, the dot GS WFs
undergo profound modifications, crossing over from weakly correlated
states at small $\lambda$ to Wigner molecular states, characterized by
strong correlations among electrons, for higher
$\lambda$~\cite{egger1,gattobigio1,ghosal1,ghosal2,ghosal3,filinov,zeng1}. The
crossover is smooth and occurs through two
phases~\cite{ghosal1,ghosal2,ghosal3}.\\
\noindent First, radial correlations begin to develop. As a result,
ring-like structures are formed. In addition, for $N\geq6$ also
centered structures appear, with the localization of one or more
electrons in the center.\\
\noindent As $\lambda$ is increased, angular correlations begin to
develop entering the {\em incipient Wigner molecule}
regime~\cite{ghosal1,ghosal2}. Eventually, for strong interactions,
the dot WF becomes a rotating Wigner molecule, with electrons
localized around positions corresponding to those of classical charged
particles parabolically confined~\cite{bedanov,jean}. Such states are
the analogue of the Wigner crystal~\cite{wigner1} but with a finite
size. Angular correlations cannot be resolved in a rotationally
invariant system but can be characterized by studying two-body angular
correlation functions, which show an oscillatory
behaviour~\cite{ghosal1,ghosal2,ghosal3}.\\
\noindent Also the WFs calculated with PHF show a behaviour in
qualitative agreement to the above results. Let us begin to introduce
the spin-resolved one-body electron density
$\rho_{1}^{s_{z}}(\mathbf{r})$, defined for a normalized dot state
$|N,L,S,S_{\mathrm z}\rangle$ as
\begin{equation}
  \rho_{1}^{s_{z}}(\mathbf{r})=\langle N,L,S,S_{z}|\hat{\psi}^{\dagger}_{s_z}(\mathbf{r})\hat{\psi}_{s_{z}}(\mathbf{r})|N,L,S,S_{z}\rangle
\end{equation} 
and $\hat{\psi}_{s_{z}}(\mathbf{r})$ defined in~(\ref{eq:fieldop}). In this section we want to illustrate the most
relevant aspects of the transition towards the Wigner molecule and
will not enter into details about the spin structure of such
states. Therefore, we only need to consider the {\em total} charge
density, summed over the spin:
\begin{equation}
  \rho_{1}(\mathbf{r})=\sum_{s_{z}=\pm1/2}\rho_{1}^{s_{z}}(\mathbf{r})\, .\label{eq:rho1}
\end{equation}
Since the dot WFs are eigenstates of the angular momentum, introducing
polar coordinates $\mathbf{r}\to(r,\theta)$ one has
$\rho_{1}(\mathbf{r})\equiv\rho_{1}(r)$.
\begin{figure}[ht]
\begin{center}
\includegraphics[width=10cm,keepaspectratio]{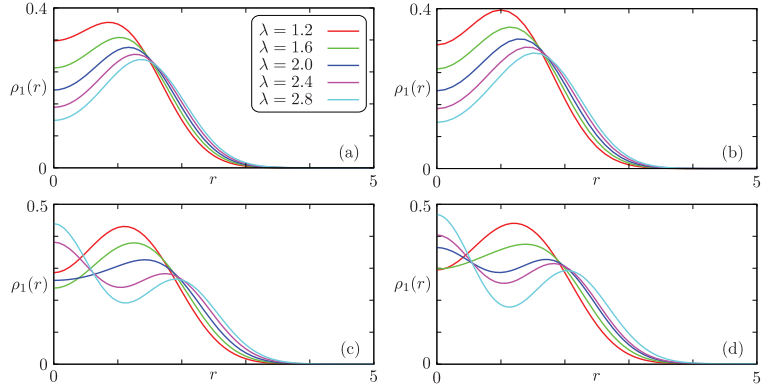}
\caption{One-body density $\rho_{1}(r)$ (units: $\ell_{0}^{-2}$) of
  the dot GS as a function of the distance from the dot center $r$,
  for different interaction strengths $\lambda$ (shown in the inset)
  and number of particles $N$: (a) $N=4$; (b) $N=5$; (c) $N=6$; (d)
  $N=7$. For dot parameters, see text. As discussed in
  Sec.~\ref{sec:dot}, here and in the following $r$ is normalized to
  $\ell_{0}$.}
\label{fig:fig2}
\end{center}
\end{figure} 
A plot of $\rho_{1}(r)$ for different values of $\lambda$ is
represented in figure~\ref{fig:fig2}. For $N=4$ and $N=5$, with
increasing interactions the density is depleted in the core of the dot
and a sharp ridge is formed at its edge, suggesting the formation of a
ring-like structure. The position of such ridge moves outwards as the
interaction strength increases. Also for $N=6,7$ a ridge develops at
the edge and moves outwards for increasing $\lambda$. Additionally,
for $N=6$ ($N=7$) the density develops a bump for $r\approx 0$ when
$\lambda\gtrsim 2$ ($\lambda\gtrsim1.8$). This behaviour is consistent
with the formation of a centered ring structure. As we shall see in
the next section, this latter rearrangement of the WF produces
detectable signatures in the transport properties. All these findings
show that, for increasing $\lambda$, radial correlations among
electrons get more pronounced.\\ 
\noindent In order to investigate the development of angular
correlations and the emergence of a Wigner molecular state, one can
introduce the two-body correlation function
\begin{equation}
  \!\!\!\!\!\!\!\!\rho_{2}^{s_{z},s_{z}'}(\mathbf{r},\mathbf{r}')=\langle N,L,S,S_{z}|\hat{\psi}^{\dagger}_{s_{z}}(\mathbf{r})\hat{\psi}^{\dagger}_{s_{z}'}(\mathbf {r}')\hat{\psi}_{s_{z}'}(\mathbf{r}')\hat{\psi}_{s_{z}}(\mathbf{r})|N,L,S,S_{z}\rangle\, ,\label{eq:rho2}
\end{equation} 
connected to the pair distribution function
$g_{s_{z},s_{z}'}(\mathbf{r},\mathbf{r}')$ by
$\rho_{2}^{s_{z},s_{z}'}(\mathbf{r};\mathbf{r}')=\rho_{1}^{s_{z}}(\mathbf{r})\rho_{1}^{s_{z}'}(\mathbf{r
}')g_{s_{z},s_{z}'}(\mathbf{r},\mathbf{r}')$~\cite{vignale}. It is
proportional to the conditional probability of finding one electron
with spin $s_{z}$ at $\mathbf{r}$, provided that another electron with
spin $s_{z}'$ is at $\mathbf{r}'$. For the qualitative discussion in
this section, we consider the {\em total} two-body correlation
function
\begin{equation}
\rho_{2}(\mathbf{r},\mathbf{r}')=\sum_{s_{z}=\pm1/2}\sum_{s_{z}'=\pm1/2}\rho_{2}^{s_{z},s_{z}'}(\mathbf{r},\mathbf{r}')\, .
\end{equation}
An example of the spin structure of the Wigner molecules will be
discussed by employing $\rho_{2}^{s_{z},s_{z}'}(\mathbf{r})$ in
Sec.~\ref{sec:linear}, in connection with transport results.\\
\noindent A natural choice for studying
$\rho_{2}(\mathbf{r},\mathbf{r}')$ is to fix $\mathbf{r}'$ at one
point on the ridge of the one-body density:
$|\mathbf{r}'|=r_{0}(N,\lambda)$ and $\theta'=0$, where
$r_{0}(N,\lambda)$ denotes the position of the off-center maximum of
$\rho_{1}(r)$ for $N$ electrons at interaction
strength $\lambda$.\\
\begin{figure}[ht]
\begin{center}
\includegraphics[width=10cm,keepaspectratio]{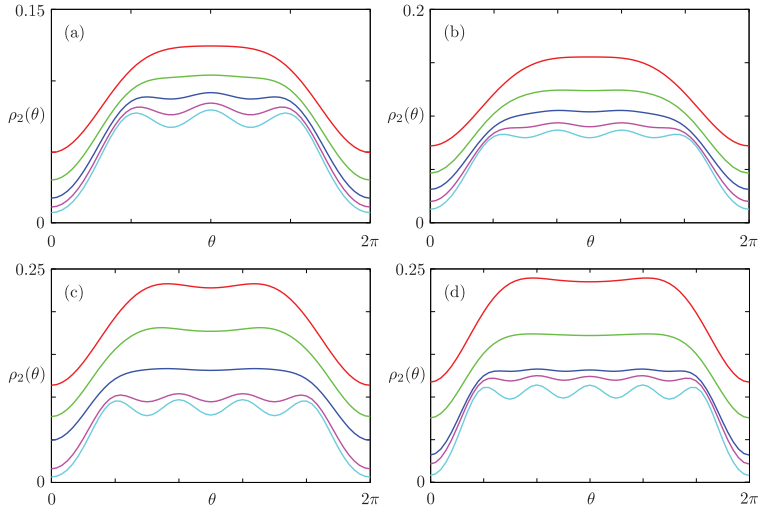}
\caption{Angular dependence of the two-body density functions
  $\rho_{2}(\theta)=\rho_{2}(r_{0}(N,\lambda),\theta,r_{0}(N,\lambda
  ),0)$ (units: $\ell_{0}^{-4}$) for the dot GSs for different values
  of $\lambda$ and (a) $N=4$; (b) $N=5$; (c) $N=6$; (d) $N=7$.
  Color-codes for $\lambda$ are the same as in
  figure~\ref{fig:fig2}.}\label{fig:fig3}
\end{center} 
\end{figure} 
\noindent Figure~\ref{fig:fig3} shows
$\rho_{2}(\theta)=\rho_{2}(r_{0}(N,\lambda),\theta,r_{0}(N,\lambda
),0)$ as a function of $\theta$ for different values of $\lambda$.
For weak interactions ($\lambda\approx 1$, red and green curves), the
correlation function is almost flat except for the ``Fermi hole'' at
$\theta=0,2\pi$, essentially induced by the Pauli exclusion
principle. This confirms that correlations among the electrons within
the ring are weak. For increasing $\lambda$, the depletion at
$\theta=0,2\pi$ gets more pronounced, signalling the increased
importance of dynamical correlations. Even more important, at the
highest values of $\lambda$ considered, $\rho_{2}(\theta)$ develops an
oscillating structure, consisting of $N-1$ maxima for states with
$N=4,5$ and with $N-2$ maxima for $N=6,7$ electrons. This is is
consistent with the discussion above, namely that angular correlations
``lag behind'' and appear for values of $\lambda$ higher than those at
which radial correlations get sizeable. Combining the information
gathered from the electron density and the two-body angular
correlation function, one can expect that for strong interactions the
dot WFs for $N=4$ and $N=5$ have the structure of a square and a
pentagon, respectively. For $N=6$ and $N=7$, they resemble a centered
pentagon and a centered hexagon.\\
\begin{figure}[ht]
\begin{center}
\includegraphics[width=10cm,keepaspectratio]{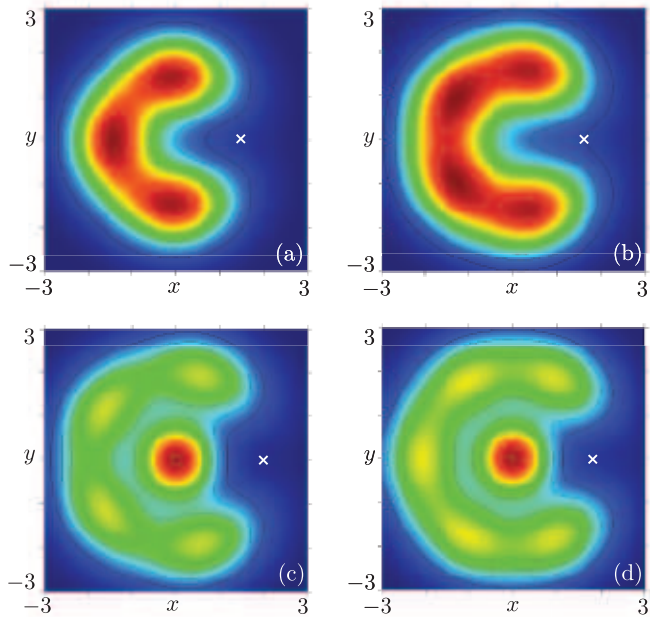} 
\caption{Density plot of the two-body density $\rho_{2}({\bf r},{\bf
    r}')$ (arbitrary units, topographic color-scale: red high and blue
    low values) for the dot GS with (a) $N=4$; (b) $N=5$; (c) $N=6$;
    (d) $N=7$ and $\lambda=2.8$. The coordinate ${\bf r}'$ of the
    probe electron is chosen on the external ridge of $\rho_1(r)$ and
    is marked with a white cross in each panel. See text for other
    physical parameters.}
\label{fig:fig4}
\end{center} 
\end{figure} 
\noindent This is confirmed by figure~\ref{fig:fig4}, which shows a
density plot of $\rho_{2}(\mathbf{r},\mathbf{r}')$ for the dot GSs as
a function of $\mathbf{r}$ in the $(x,y)$ plane. The white cross
denotes the position of $\mathbf{r}'$, which is the same as in
figure~\ref{fig:fig3}. Around $\mathbf{r}'$, the presence of the Fermi
hole is clear. Strong radial and angular correlations are observed,
confirming the structures for $N=4$ (square), $N=5$ (pentagon), $N=6$
(centered pentagon) and $N=7$ (centered hexagon).
\begin{figure}[h]
\begin{center}
\includegraphics[width=10cm,keepaspectratio]{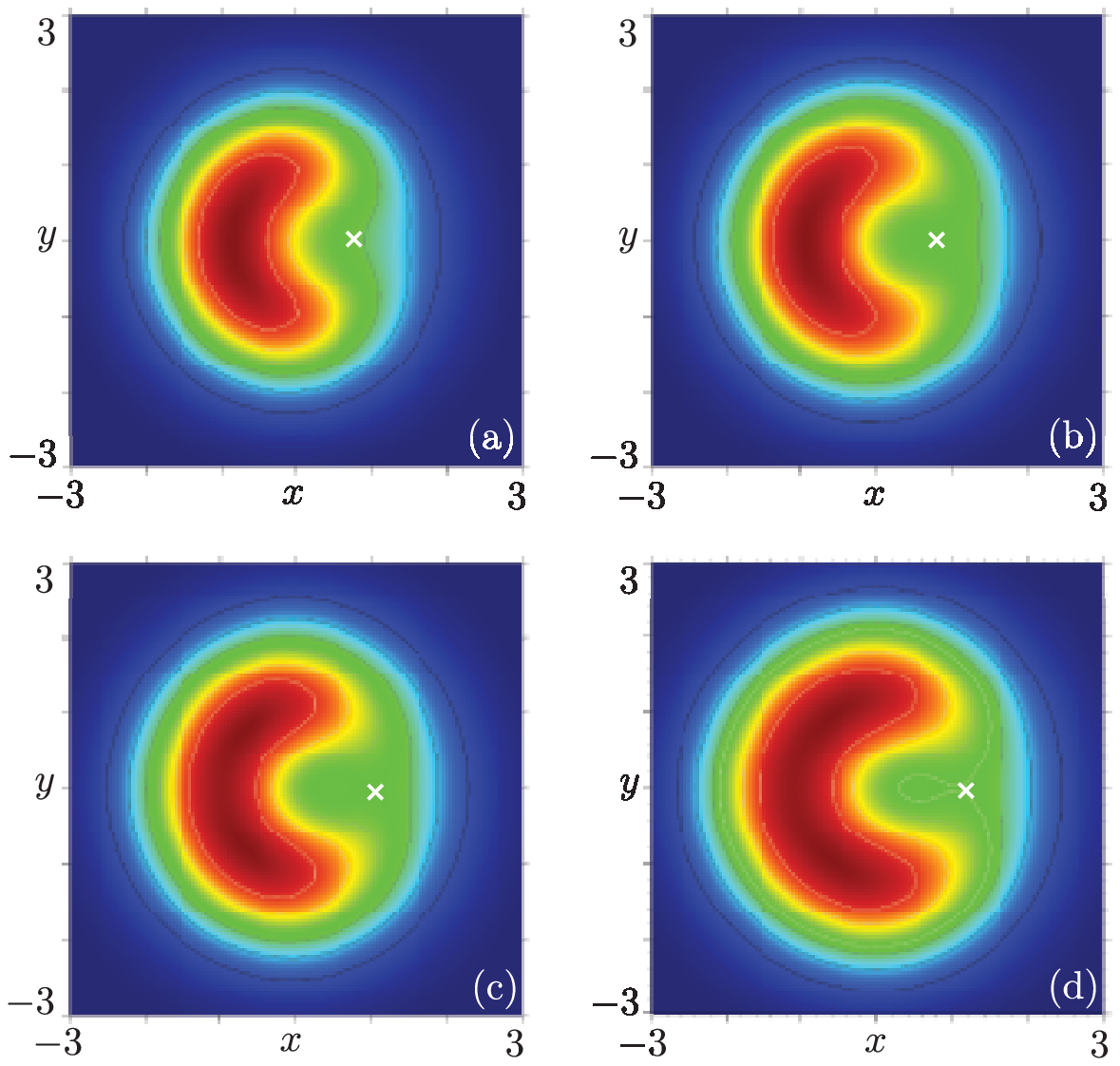} 
\caption{Same as in figure~\ref{fig:fig4} but with $\lambda=1.2$ in all
  panels.}
\label{fig:fig5}
\end{center} 
\end{figure} 
For comparison, density plots for $\lambda=1.2$ are shown in
figure~\ref{fig:fig5}. For such a smaller interaction, radial and
angular correlations beyond the Fermi hole are undetectable. The
situation is more reminiscent of a liquid-like behaviour.\\
\noindent Even though the discussion has been focused on the dot
ground state, also the WFs for the excited states behave in a similar
manner.\\ The above results confirm that PHF is able to capture at the
qualitative level all the relevant correlations of the dot WFs and to
produce Wigner molecular states. In this respect, we note that the
onset in $\lambda$ for the development of strong radial and angular
correlations in the WFs predicted by the PHF method seems to be
smaller than the one found with other techniques. As an example, for
$N=6$, both exact diagonalization~\cite{maximaging} and density
functional~\cite{gattobigio1} calculations predict the localization of
one electron in the dot center for $\lambda\approx8$ while from the
PHF calculations one would obtain $\lambda\approx2$. A similar
tendency to underestimate the crossover in $\lambda$ for the
transition between different dot GSs has already been observed in
earlier studies of PHF~\cite{PHF}. Since the {\em qualitative} changes
of the dot WF are correctly captured by PHF, we expect that the
transport results described below will be at least qualitatively
correct.
\subsection{Transport properties} 
In this section we will show how modifications of the WF, occurring in
the transition from a liquid to a molecular character, can be detected
in the transport properties. In the rest of the paper, we assume {\em
  symmetric} tunnelling barriers, with
$\Gamma^{(\mathrm{E})}=\Gamma^{(\mathrm{C})}=\Gamma_{0}$ in~(\ref{eq:ratepil}). Typical values for $\Gamma_{0}$ are of the
order of some MHz.
\subsubsection{Linear transport} 
\label{sec:linear}
We start considering the linear regime ($V\to 0$), which provides
information on the dot ground states.
\begin{figure}[ht]
\begin{center}
  \includegraphics[width=10cm,keepaspectratio]{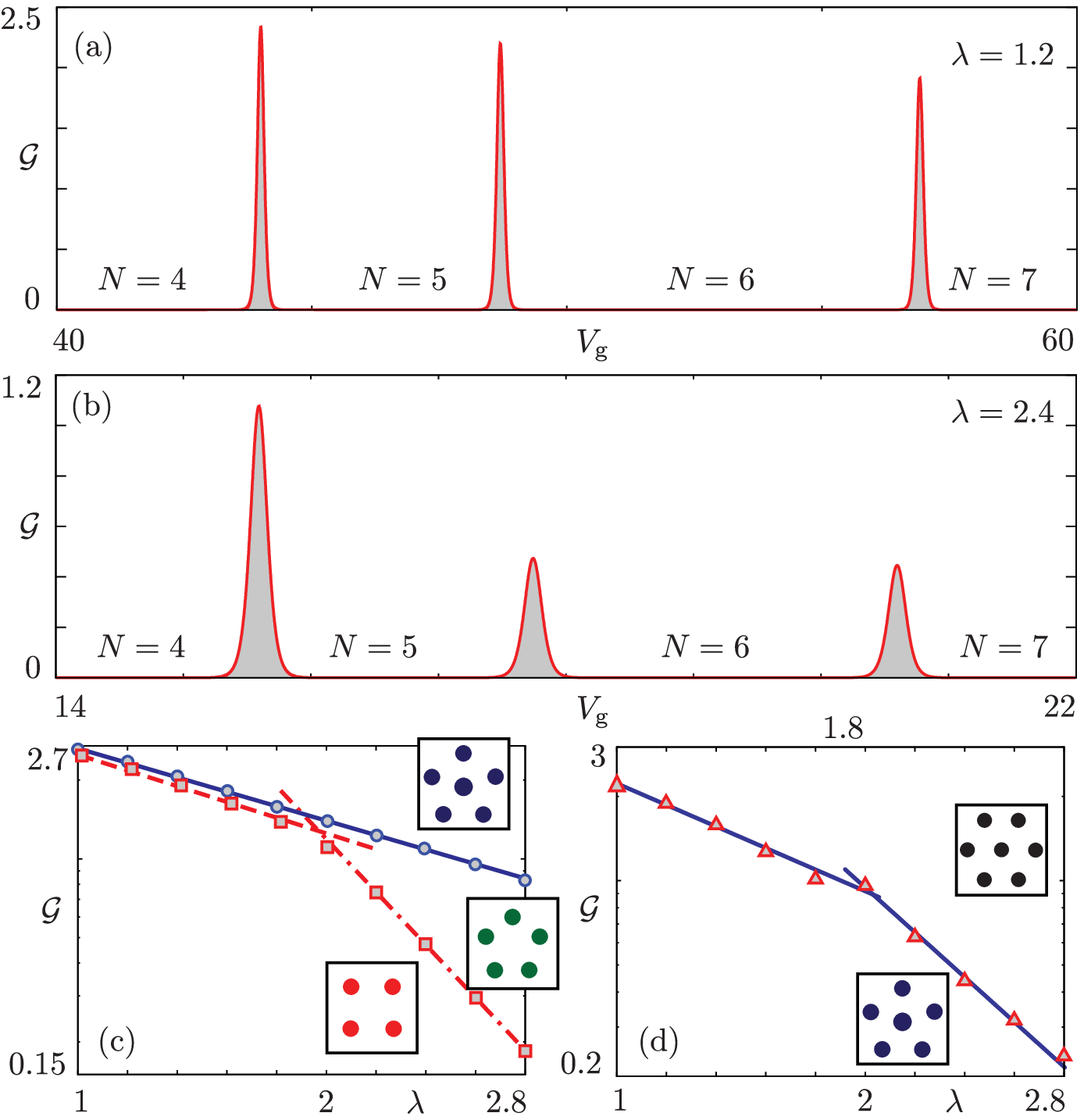} 
  \caption{(a) Plot of the linear conductance $\mathcal G$ as a
    function of $V_{\mathrm g}$ (units mV) for $T=500$ mK and
    $\lambda=1.2$.  (b) Same as in (a) but for $\lambda=2.4$. (c) Plot
    of the conductance maxima (log scale) as a function of the
    interaction strength $\lambda$ for the transition
    $4\leftrightarrow5$ (circles) and $5\leftrightarrow6$
    (squares). Lines are a guide for the eye. (d) Same as in (c) but
    for the transition $6\leftrightarrow7$. The insets in panels (c,d)
    represent the structure of the dot GS WF in the strongly
    correlated regime. Conductance unit
    $\mathcal{G}_{0}=1.6\cdot10^{-7}\Gamma_{0}$
    s/G$\Omega$}. \label{fig:fig7}
\end{center} 
\end{figure} 
A plot of the linear conductance ${\mathcal G}$ as a function of
$V_{\mathrm g}$, calculated solving numerically~(\ref{eq:rateq}) in
the stationary regime is shown in figure~\ref{fig:fig7}(a). It has
been calculated for $\lambda=1.2$. The conductance exhibits the well
known Coulomb oscillations: conductance peaks are separated by regions
where the dot is in the Coulomb blockade regime and transport is
forbidden~\cite{kouw2,beenakker}. Peaks occur when the chemical
potential of the dot is aligned with the electrochemical potential of
the leads $\mu_{\mathrm D}=\mu_{0}$, which is satisfied for a given
transition $N\leftrightarrow N+1$ by suitably tuning $V_{\mathrm
  g}$. Since $\mu_{0}$ simply induces a constant shift of the position
of the linear conductance peaks in $V_{\mathrm g}$, we assume
$\mu_{0}=0$. Turning to stronger interactions $\lambda=2.4$,
figure~\ref{fig:fig7}(b), the linear conductance decreases. The
observed suppression of $\mathcal{G}$ as $\lambda$ is increased can be
interpreted as due to the increased difficulty to tunnel into (or out
from) an electronic system with strong Coulomb repulsion. However, the
conductance peaks for the transition $5\leftrightarrow6$ and
$6\leftrightarrow7$ have been suppressed much more than that corresponding to $4\leftrightarrow 5$.

\noindent In order to investigate this behaviour more systematically,
the heights of the conductance peaks are shown in logarithmic scale as
a function of $\lambda$ in figure~\ref{fig:fig7}(c,d). For the
transition $4\leftrightarrow5$ (circles), a single slope is
observed, signalling an exponential suppression of the conductance as
$\lambda$ increases. On the other hand, for $5\leftrightarrow6$
(squares) a bimodal behaviour occurs, with a slope for $\lambda\leq 2$
that is very similar to the one found for $4\leftrightarrow5$. A
steeper slope is found for $\lambda>2$. The conductance peak for the
transition $6\leftrightarrow7$, see figure~\ref{fig:fig7}(d), shows a
behaviour similar to $5\leftrightarrow6$: a smaller slope for
$\lambda\leq 1.6$ and a steeper one for $\lambda\geq 2$.\\
\begin{figure}[ht]
\begin{center}
  \includegraphics[width=10cm,keepaspectratio]{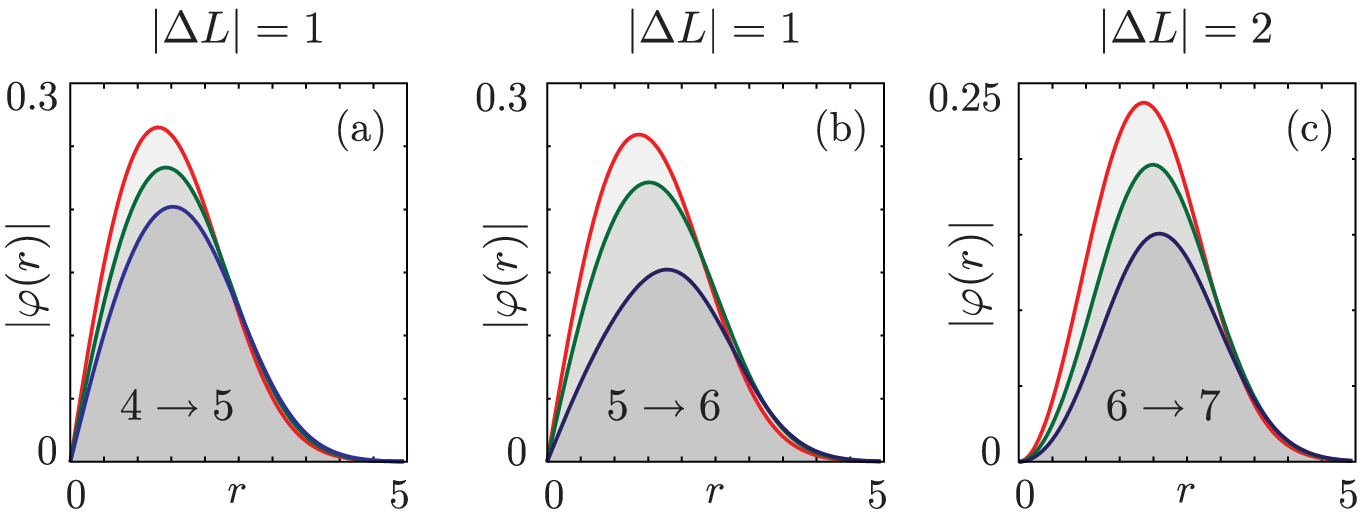}
  \caption{Radial behaviour of the modulus of the QPWF $|\varphi(r)|$
  -- see~(\ref{eq:qpwfs}) -- for $\lambda=1.2$ (red),
  $\lambda=1.8$ (green) and $\lambda=2.4$ (blue) and different
  transitions between dot GSs: (a) $4\leftrightarrow 5$; (b)
  $5\leftrightarrow 6$; (c) $6\leftrightarrow 7$. At the top of
  the panels, $|\Delta L|$ for the given transition is quoted, see
  table~\ref{tab:tab1}.}
\label{fig:fig8}
\end{center} 
\end{figure}
\noindent In order to interpret these behaviours we can deduce more
precise information about the tunnelling of electrons from the QPWF,
see~(\ref{eq:qpwfs}). Figure~\ref{fig:fig8} shows its modulus
$|\varphi(r)|$ for the transition between dot GSs $N\to N+1$ with
$N=4$ (a), $N=5$ (b) and $N=6$ (c) and increasing values of
$\lambda$. Since all these transitions have $|\Delta L|\neq 0$ (see
table~\ref{tab:tab1}), the WF exhibits an off-center maximum and is small around
the origin, hence tunnelling is strongly suppressed in the center while
it is enhanced at the edge of the dot. The above transport results are
now explained by considering both the shape of the QPWF and the
structure of the WF of the dot GS for each $N$, discussed in
Sec.~\ref{sec:molecules}.\\
\noindent On the one hand, by comparing the WFs for two subsequent
dot GSs one can estimate where the tunnelling electron should enter in
order to provide an optimal matching between the dot states and obtain
a good transmission through the dot. On the other hand, the most
likely position of the tunnelling electron is essentially dictated by
$|\Delta L|$, as the QPWF shows. As a result, a higher conductance is
obtained in situations where the QPWF is peaked so as to provide a
maximal overlap of the dot WFs. With these considerations, let us
now reexamine figure~\ref{fig:fig7}(c,d).\\
\noindent For the transition $4\leftrightarrow 5$, as $\lambda$
increases, the dot WFs build up radial and subsequently angular
correlations, ending up eventually in a molecular state with a square
($N=4$) or pentagon ($N=5$) symmetry with always a ring-like
structure. As such, maximum overlap is achieved when the
tunnelling electron jumps to the edge of the dot. This is the case, in
agreement with the results of the QPWF, as confirmed by
figure~\ref{fig:fig8}(a).\\
\noindent The transition $5\leftrightarrow 6$ displays a more
interesting double-slope feature. For small $\lambda$, both the dot GS
WFs for $N=5$ and $N=6$ display weak correlations and have a ring-like
structure. Similar to the case discussed above the tunnelling
electron, entering at the edge of the dot, provides an optimal overlap
of the dot WFs. Therefore, a slope similar to the one observed for
$4\leftrightarrow 5$ is obtained for small $\lambda$. On the other
hand, for $\lambda\gtrsim 2$ one electron is shifted towards the
center of the dot. Eventually, the WF for $N=6$ acquires the shape of
a centered pentagon, see the inset in figure~\ref{fig:fig7}(c). The
optimal overlap would be achieved with the tunnelling electron jumping
to the center of the dot. This however is not allowed for dynamical
reasons, as shown by the QPWF in figure~\ref{fig:fig8}(b): the
tunnelling electron needs to enter into the dot edge. Therefore, an
{\em additional} suppression of the conductance occurs, which is
detected in the sharp change of slope of $\mathcal{G}$ shown in
figure~\ref{fig:fig7}(c).\\
\noindent In the case of $6\leftrightarrow7$, for $\lambda<1.8$
correlations in both the WFs are weak and the latter exhibit a
ring-like shape as in all the low $\lambda$ regimes already
discussed. For $\lambda\approx 1.8$, the GS with $N=7$ begins to shift
one electron towards the center of the dot, while the GS of $N=6$
remains annular. Since the QPWF is peaked at the edge of the dot, this
corresponds to a slight, yet noticeable, suppression of $\mathcal{G}$
visible in figure~\ref{fig:fig7}(d) for $\lambda=1.8$. For
$\lambda\geq2$, also the GS for $N=6$ has one electron in the center
of the dot. The optimum overlap is again achieved for tunnelling at
the dot edge and therefore one could expect a return of a power law
similar to the one observed for $\lambda\leq1.6$. On the contrary, one
observes a steeper slope. In order to explain this phenomenon we need
to consider in detail the spin structure of the dot
GSs.\\ 

\noindent We consider here the two states
$|N,L,S,S_{z}\rangle=|6,0,0,0\rangle$ and $|7,2,1/2,1/2\rangle$.
\begin{figure}[h]
\begin{center}
  \includegraphics[width=12cm,keepaspectratio]{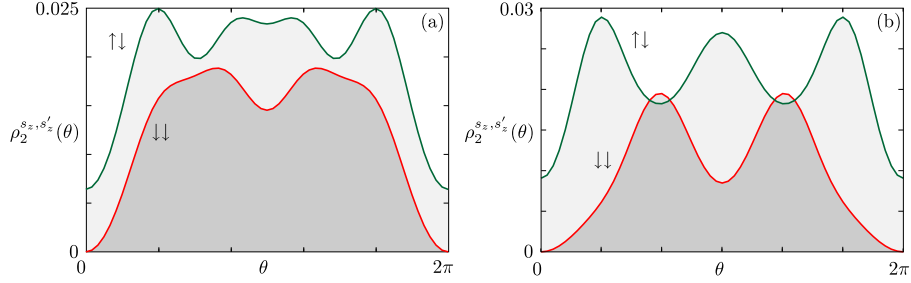}
  \caption{Angular dependence of the spin-resolved two-body density
    functions
    $\rho_{2}^{s_{z},s_{z}'}(\theta)=\rho_{2}^{s_{z},s_{z}'}(r_{0}(N,\lambda),\theta,r_{0}(N,\lambda
    ),0)$ (units: $\ell_{0}^{-4}$) with $s_{z}'=-1/2$ and $s_{z}=-1/2$
    (red line) or $s_{z}=+1/2$ (green line) for the dot states (a)
    $|6,0,0,0\rangle$; (b) $|7,2,1/2,1/2\rangle$. Here,
    $\lambda=2.4$.}
\label{fig:fig8b}
\end{center} 
\end{figure} 
Figure~\ref{fig:fig8b} shows the spin-resolved two body correlation
function $\rho_{2}^{s_{z},s_{z}'}(\mathbf{r},\mathbf{r}')$
in~(\ref{eq:rho2}), calculated along the outer ring of the quantum dot
GSs for $N=6,7$ and a representative value $\lambda=2.4$. One electron
with spin down is assumed to lie at $r'=r_{0}(N,\lambda)$ and
$\theta'=0$. Starting with $N=6$, panel (a), one can see that the
probability of finding another spin-down electron
($\rho_{2}^{\downarrow,\downarrow}(\theta)$) is peaked around
$\theta\approx4\pi/5$ and $\theta\approx6\pi/5$, while the one for an
electron with spin up ($\rho_{2}^{\uparrow,\downarrow}(\theta)$) is
highest at $\theta=2\pi/5$ and $\theta=8\pi/5$, although two relative
maxima for this probability are observed also for $\theta=4\pi/5$ and
$\theta=6\pi/5$. For $N=7$ one finds that the probability of finding
another spin-down electron
($\rho_{2}^{\downarrow,\downarrow}(\theta)$) is largest at
$\theta=2\pi/3$ and $\theta=4\pi/3$, while spin-up electrons are
maximally likely at $\theta=\pi/3,\pi,5\pi/3$.\\
\begin{figure}[h]
\begin{center}
  \includegraphics[width=10cm,keepaspectratio]{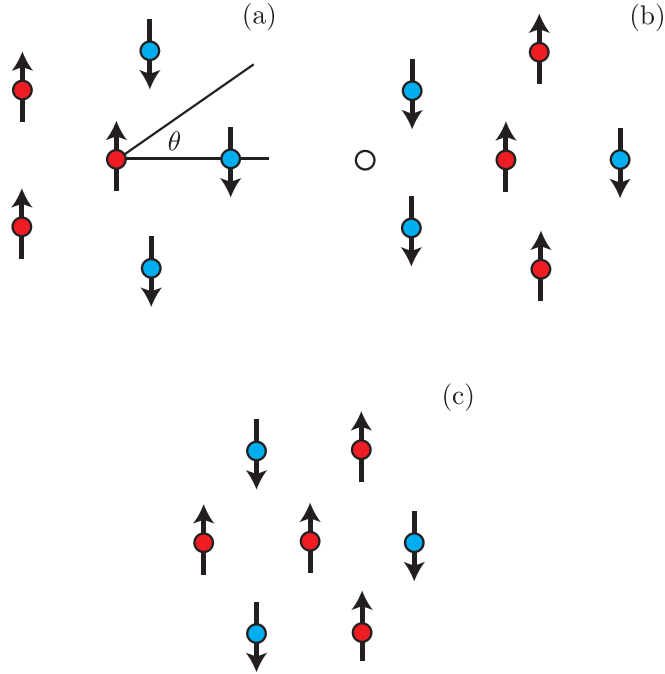}
  \caption{The possible spin pattens, most relevant for linear
    transport, contributing to the GS WF for $\lambda=2.4$ and (a,b)
    $N=6$; (c) $N=7$. The probe electron lies at $\theta=0$ and the
    white dot denotes the preferred site for tunnelling.}
\label{fig:fig8c}
\end{center} 
\end{figure} 

\noindent In the Wigner molecule regime, only a few different spin
configurations may contribute to the dot WF. In the case of $N=6$,
once the spin direction for the electron at the center of the dot is
chosen, only two possible spin arrangements are possible for the
pentagon at the edge. These are shown in Figs.~\ref{fig:fig8c}(a,b)
for the case of a spin up in the dot center. Since the WF for $N=6$ is
a spin singlet, also the two other configurations, obtained flipping
all spins in panels (a) and (b) are possible (not shown). Since the
correlation function in figure~\ref{fig:fig8b}(a) for parallel
spin-down electrons is more peaked around $\theta=4\pi/5$ and
$\theta=6\pi/5$, one can anticipate that the configuration represented
in panel (b) contributes more than the one in panel (a). Also for
$N=7$ several spin configurations for the dot edge exist, once the
spin in the center of the dot has been fixed. However, the clear peak
structure of figure~\ref{fig:fig8b}(b) strongly suggests a
well-defined texture of {\em alternating spins} in the outer ring of
the molecule with a corresponding spin-up electron in the center of
the dot, consistent with $S_{z}=1/2$. Therefore, we can infer that the
dot WF for $N=7$ has the spin structure shown in
figure~\ref{fig:fig8b}(c). Note that the discussion for the other
states of the multiplet for $N=7$ is identical, provided that one
flips all spins for the states with $S_{z}=-1/2$.\\ Let us go back to
the transport properties of the dot. Among all the possible spin
configurations for $N=6$, the ones with a spin down at the dot center,
obtained by flipping the spins of those shown in
figure~\ref{fig:fig8c}(a,b), provide a very poor overlap with the
state with $N=7$ and therefore can be neglected. The configuration
shown in figure~\ref{fig:fig8c}(a) also provides a negligible overlap:
there is no position around the edge for the tunnelling electron so
that the final state has the same spin pattern shown in
figure~\ref{fig:fig8c}(c). Concerning the situation shown in
Fig~\ref{fig:fig8c}(b), the tunnelling electron can only jump in the
proximity of $\theta=\pi$ (white dot in the figure) since all other
positions would lead to a wrong spin pattern on the edge. This results
in a suppression of the tunnelling amplitude as compared to the case
of small $\lambda<1.8$, when the tunnelling electron is free to
delocalize around the ring due to the negligible correlations of the
dot WF.\\ It is important to note that the situation described before
does not occur either for $4\leftrightarrow 5$, whose conductance is
featureless, or for $5\leftrightarrow 6$ whose change of slope is
mainly related to the localization of one
electron in the center of the dot.\\

\noindent From the above discussions one can conclude that the
transition towards the Wigner molecule, accompanied by qualitative
rearrangements of the charge or spin textures of the dot WF, may be
detectable in the linear transport properties. This seems particularly
relevant when transport involves states with higher numbers of
electrons and intricate spin patterns, such as $N=6,7$, due to the
complex internal structure. Simpler configurations such as the ones
for the transition $4\leftrightarrow5$ discussed above may not cause
any signature in transport. It is worth to notice that the spin
effects discussed above are subtler than the more common type-II spin
blockade~\cite{wein1,wein2}. In the latter, the current flow is
blocked due to the impossibility to fulfil total spin conservation
by tunneling events. In the Wigner molecule regime, on the other hand,
even if spin conservation is satisfied an {\em additional suppression}
of the current as $\lambda$ increases occurs, due to the peculiar
internal spin structure of the dot
WFs.\\
 
\subsubsection{Nonlinear transport}
In the nonlinear regime, transport also triggers the population of
excited states of the quantum dot. In this section we discuss one
particular case, to show signatures of the transition towards the
Wigner molecule. To be specific, we will concentrate on the regime
where only states with $N=5,6$ electrons in the dot are
involved. Furthermore, we assume strong relaxation: $W\gg\Gamma_{0}$,
see~(\ref{eq:relaxation}).\\
\begin{figure}[ht]
\begin{center}
\includegraphics[width=10cm,keepaspectratio]{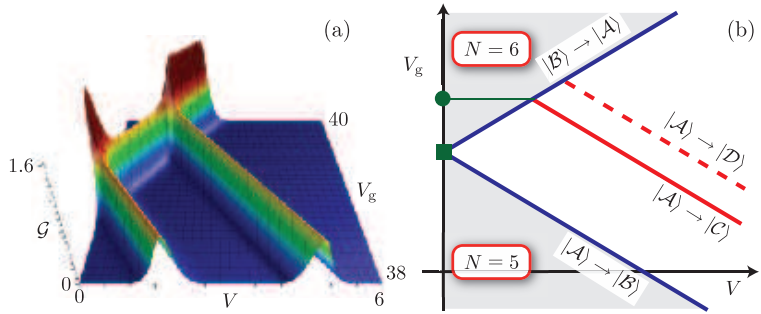} 
\caption{(a) Nonlinear conductance $\mathcal{G}$ (units
  $\mathcal{G}_{0}$ same as in figure~\ref{fig:fig7}) as a function of
  applied voltage $V$ (units mV) and gate voltage $V_{\mathrm g}$
  (units mV) for $\lambda=1.4$, $T=750$ mK and $W=50\Gamma_{0}$. (b)
  Scheme of the transitions involving the ground states for $N=5,6$
  and the lowest lying excited states for $N=6$. The dashed line
  represents a spin-blockaded transition not visible in the sequential
  regime.}
\label{fig:figs}
\end{center} 
\end{figure} 

\begin{table}[h]
  \caption{\label{tab:tab2} States and corresponding quantum numbers involved in transport dynamics for the range of $V$ and $V_{\mathrm g}$
    considered in figure~\ref{fig:figs}.}
\begin{indented}
\item[]\begin{tabular}{@{}lllll} 
\br
State&N&L&S&$S_{z}$\\ 
\mr 
$|{\mathcal A}\rangle$&5&$\pm1$&1/2&$\pm1/2$\\ 
$|{\mathcal B}\rangle$&6&0&0&0\\
$|{\mathcal C}\rangle$&6&$\pm 1$&1&0,$\pm 1$\\ 
$|{\mathcal D}\rangle$&6&0&2&0,$\pm1$,$\pm2$\\
\br 
\end{tabular} 
\end{indented} 
\end{table} 
\noindent The numerically evaluated $\mathcal{G}$ as a function of $V$
and $V_{\mathrm g}$ is shown in figure~\ref{fig:figs}(a) for
$\lambda=1.4$. It exhibits conductance lines corresponding to
transitions between the dot GSs or between GS and lowest-lying excited
states. The scheme of the expected lines is shown in
figure~\ref{fig:figs}(b) for the voltages region considered here. The
transitions corresponding to each line are shown, with their quantum
numbers given in table~\ref{tab:tab2}. The blue lines represent
transitions between the dot GSs for $N=5$ ($|{\mathcal A}\rangle$) and
$N=6$ ($|{\mathcal B}\rangle$). The red lines represent channels
involving the GS of $N=5$ ($|{\mathcal A\rangle}$) and one of the
first two excited multiplets of $N=6$: the lowest one is denoted as
$|{\mathcal C}\rangle$ while the next-to-lowest is $|{\mathcal
  D}\rangle$. Since calculations are performed for temperatures
smaller than the average level spacing between the dot multiplets, in
the strong relaxation regime transitions among the excited states of
the dot cannot occur. Each of these transition lines corresponds to
the opening of the specific transport channel involving an excited
state of the dot with $N=6$. Note that transitions involving excited
states for $N=5$ are not present in the considered range of $V$ and
$V_{\mathrm g}$, since they lie at higher energies. A comparison
between the scheme of figure~\ref{fig:figs}(b) and the calculated
conductance, figure~\ref{fig:figs}(a) shows that only the first
transition line (red solid), corresponding to $|{\mathcal
  A}\rangle\to|{\mathcal C}\rangle$ is observed, while the one
corresponding to the second excited multiplet of $N=6$ (red dashed) is
absent. By inspecting table~\ref{tab:tab2}, one notes that the
transition $|{\mathcal A}\rangle\to|{\mathcal D}\rangle$ involves
$|\Delta S|>1/2$ and therefore is forbidden~\cite{wein1,wein2},
leading to a vanishing conductance.\\
\begin{figure}[ht]
\begin{center}
  \includegraphics[width=10cm,keepaspectratio]{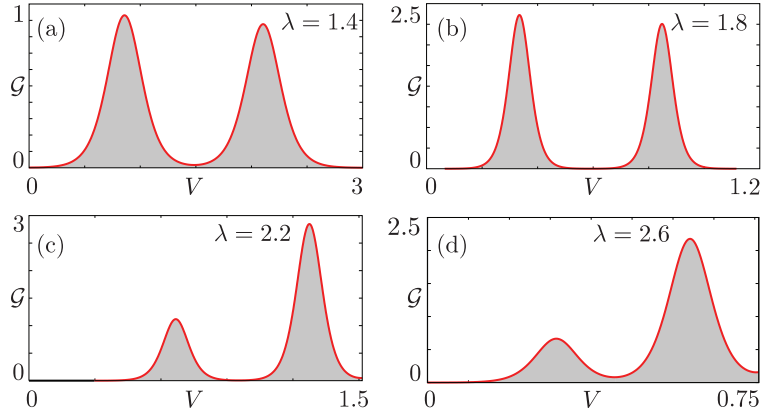} 
\caption{Nonlinear
    conductance as a function of the applied voltage $V$ (units mV)
    near the transition $5\leftrightarrow6$ for (a) $\lambda=1.4$,
    $V_{\mathrm g}=39.3$ mV, $T=750$ mK; (b) $\lambda=1.8$,
    $V_{\mathrm g}=27.3$ mV, $T=150$ mK; (c) $\lambda=2.2$,
    $V_{\mathrm g}=20.2$ mV, $T=150$ mK; (d) $\lambda=2.6$,
    $V_{\mathrm g}=15.9$ mV, $T=100$ mK. In all plots,
    $W=50\Gamma_{0}$ and the conductance unit $\mathcal{G}_{0}$ is the
    same as in figure~\ref{fig:fig7}.}
\label{fig:fig9}
\end{center} 
\end{figure} 

\noindent Figure~\ref{fig:fig9} shows plots of $\mathcal{G}$ as a
function of the applied voltage for different values of $\lambda$. In
all panels, the value of $V_{\mathrm g}$ has been chosen to lie
between the green square and the green dot in
figure~\ref{fig:figs}(b). The peak at lower $V$ in
Figs.~\ref{fig:fig9}(a--d) corresponds to the GS to GS transition
$|{\mathcal A}\rangle\to|{\mathcal B}\rangle$, while the second one to
the transition $|{\mathcal A}\rangle\to|{\mathcal C}\rangle$. As is
visible from the voltage ranges of the plots, the dot level spacing
(corresponding to the distance between the nonlinear conductance
peaks) gets narrower as $\lambda$ is increased. In order to be able to
resolve both conductance peaks, calculations for higher $\lambda$ have
been performed at lower temperatures than those at smaller
$\lambda$.\\
\noindent Comparing the panels (a) and (b) for
$\lambda<2$ with panels at $\lambda>2$ (c) and (d), a qualitative
difference in the behaviour is easily observed: for weaker interaction
strengths, the first peak is always higher than the second one, while
for $\lambda>2$ the situation reverses drastically.\\
\noindent Such a behaviour {\em cannot} be attributed to the
difference in temperature between different calculations. Indeed,
calculations for $\lambda=1.4$ and $\lambda=1.8$ performed at {\em
lower} temperatures display narrower conductance peaks but still with
almost equal height. Increasing $T$ for $\lambda=2.2$ and
$\lambda=2.6$ always suppresses the conductance for the GS to GS peak
with respect to the transition towards the excited state in all the
temperature range in which the two peaks are resolved.\\
\begin{figure}[ht]
\begin{center}
\includegraphics[width=12cm,keepaspectratio]{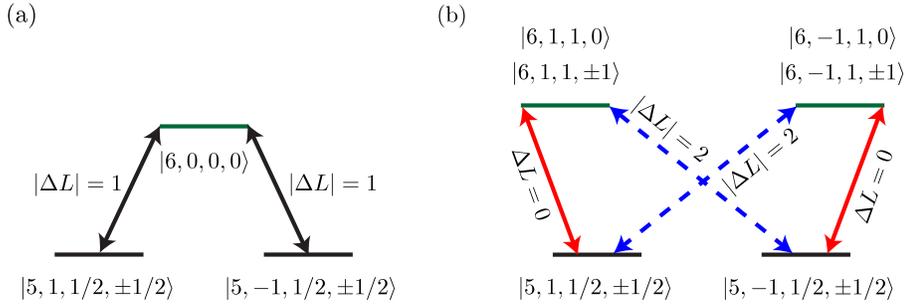} 
\caption{(a) Scheme of the transitions involved in the first
  conductance peak shown in figure~\ref{fig:fig9}. (b) Same as in (a)
  but for the second conductance peak.}
\label{fig:fig10pre}
\end{center} 
\end{figure} 
\noindent The behaviour of the nonlinear conductance can be related to
qualitative changes in the WFs of the excited states for
$N=6$. Figure~\ref{fig:fig10pre} shows the states involved in the
transport dynamics of the two conductance peaks discussed above. The
height of each peak is determined by the available transport channels
and by the transition rates connecting the dot states. As the
interaction strength is increased the first conductance peak,
involving only transitions between GSs -- see
figure~\ref{fig:fig10pre}(a) -- behaves exactly as the linear
conductance peak discussed in Sec.~\ref{sec:linear}. The height of the
second peak is on the other hand determined by two families of
transport channels connecting the GS with $N=5$ to the excited
multiplet of $N=6$: as shown in figure~\ref{fig:fig10pre}(b), channels
with either $\Delta L=0$ or $|\Delta L|=2$ are possible. The
corresponding transition amplitudes are shown in
figure~\ref{fig:fig10}(a) as a function of $\lambda$. Transition rates
are proportional to these amplitudes, see~(\ref{eq:ratepil}). One can
see that for $\lambda<2$ the transition channel with $\Delta L=0$ is
strongly suppressed, while the one with $|\Delta L|=2$ is larger and
decaying with $\lambda$. For $\lambda>2$, a sudden decrease of the
transition amplitude for the channel with $|\Delta L|=2$ is found,
while the one for the channel with $\Delta L=0$ jumps to a very large
value.\\
\begin{figure}[ht]
\begin{center}
  \includegraphics[width=13cm,keepaspectratio]{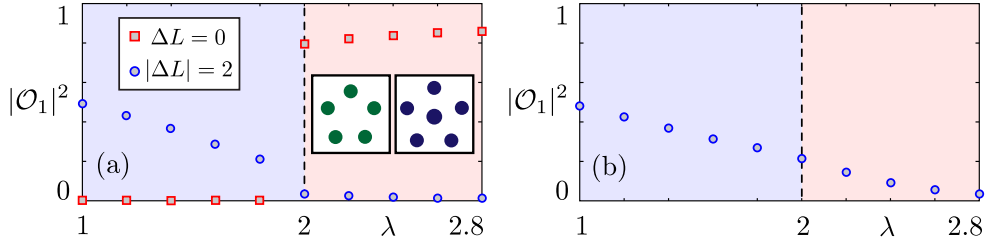}
  \caption{(a) Squared modulus of the transition amplitude
    $|\mathcal{O}_{1}|^2$ -- see~(\ref{eq:amplitudes}) -- for the
    processes $|5,\pm1,1/2,\pm1/2\rangle\to|6,\mp1,1,\pm 1\rangle$
    with $|\Delta L|=2$ (circles) and
    $|5,\pm1,1/2,\pm1/2\rangle\to|6,\pm1,1,\pm 1\rangle$ with $\Delta
    L=0$ (squares) shown in figure~\ref{fig:fig10pre}(b). The insets
    show the character of the dot WF for high values of $\lambda$. (b)
    Same as in (a) but for the GS to GS transition
    $|5,\pm1,1/2,\pm1/2\rangle\to|6,0,0,0\rangle$ shown in
    figure~\ref{fig:fig10pre}(a). Amplitudes for transitions involving
    triplet states with $S_{z}=0$ are proportional to the ones shown
    here, the proportionality constant being a Clebsch-Gordan factor
    $1/\sqrt{2}$. Amplitudes for transitions with $|\Delta S_{z}|>1/2$
    are zero.}
\label{fig:fig10}
\end{center} 
\end{figure} 
\noindent This peculiar behaviour can be again explained using the
QPWFs.
\begin{figure}[ht]
\begin{center}
  \includegraphics[width=10cm,keepaspectratio]{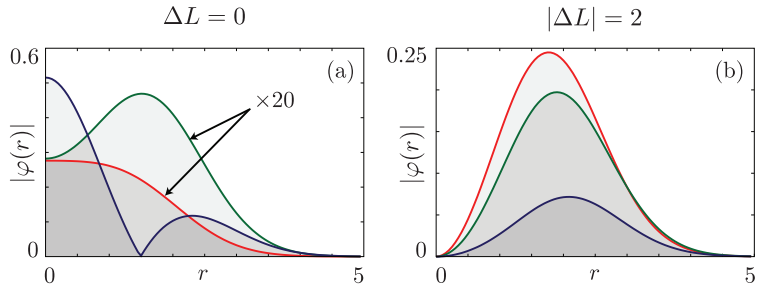} 
  \caption{Radial behaviour of the modulus of the QPWF $|\varphi(r)|$
    for $\lambda=1.2$ (red) $\lambda=1.8$ (green) and $\lambda=2.4$
    (blue) and different processes: (a) $5\leftrightarrow 6$ GS to
    first excited with $\Delta L=0$, i.e. $|5,\pm
    1,1/2,\pm1/2\rangle\leftrightarrow|6,\pm 1,1,\pm1\rangle$; (b)
    $5\leftrightarrow 6$ GS to first excited with $|\Delta L|=2$,
    i.e. $|5,\pm 1,1/2,\pm1/2\rangle\leftrightarrow|6,\mp
    1,1,\pm1\rangle$. Note that in (a), the red and green curves for
    $\lambda=1.2$ and $\lambda=1.8$ have been magnified by a factor 20.}
\label{fig:fig11}
\end{center} 
\end{figure} 
Figure~\ref{fig:fig11} shows the QPWF for the transition with $\Delta
L=0$ (panel a) and for the one with $\Delta L=2$ (panel b). When
$\lambda<2$, the case of $\Delta L=0$ has nonzero amplitude near the
center of the dot and consequently a very small overlap between the
configurations for $N=5$ and $N=6$ which have a ring-like
character. This results in a very small transition amplitude. On the
other hand, the case of $|\Delta L|=2$ has a large QPWF near the edge
of the dot and therefore provides a much better overlap between the
dot configurations. The situation reverses dramatically for $\lambda>
2$: the developing radial correlations induce a qualitative change in
the dot WF for $N=6$ and one electron moves near the center of the
dot, in analogy to the case of the GS. In this case, a much better
overlap occurs for $\Delta L=0$, since the extra electron preferably
sits in the center of the dot, see the right insets in
figure~\ref{fig:fig10}(a). The transition with $|\Delta L|=2$ is
strongly suppressed since the corresponding QPWF is $\propto r^{2}$
and thus negligible near the dot center.\\
\noindent Therefore, for $\lambda< 2$ transport through the excited
state occurs essentially via the channel with $|\Delta L|=2$, whose
amplitude is similar to the one for the GS to GS transition, see
figure~\ref{fig:fig10}(b). This explains why, for $\lambda< 2$, the
two conductance peaks in figure~\ref{fig:fig9}(a,b) have almost equal
height. For $\lambda>2$ the channel with $\Delta L=0$ clearly
dominates, due the peculiar rearrangement of the dot WF for $N=6$
moving towards the Wigner molecule. The amplitude for this channel is
larger than the one for the GS to GS transition and therefore the
conductance peak for the transition involving the excited state is
higher than the one for the GS to GS transition as shown in
figure~\ref{fig:fig9}(c,d).
\section{Conclusions}
\label{sec:conclusions}
In this paper we have investigated correlation effects in a quantum
dot via linear and nonlinear transport, employing the PHF
technique. For increasing interaction strength, the ground and excited
dot states have been analyzed for $4\leq N\leq 7$. As the strength of
the Coulomb interactions increases, the dot WFs build up radial and
angular correlations, smoothly crossing over from a liquid-like regime
to Wigner molecular states. Most strikingly, we have demonstrated that
signatures of such a crossover may appear both in the linear and in
the nonlinear transport properties. These signatures have been
interpreted with the systematic study of both two-body correlation
functions and
QPWFs.\\
\noindent In the linear regime, we have observed
an exponential suppression of the conductance as the transition
towards the Wigner molecule takes place. In cases when the latter is
accompanied by strong qualitative rearrangements of the dot WFs,
strong mismatches of the dot WFs involved in the transport process may
occur. This leads to a {\em stronger suppression} of the conductance,
as observed for the case $5\leftrightarrow6$. A mismatch of the dot
WFs due to the emergence of particular {\em spin} structure of the dot
states may also occur, as exemplified by the case of
$6\leftrightarrow7$. Also this fact leads to an increased suppression
of the linear conductance.\\
\noindent In the nonlinear regime, the conductance may even be {\em enhanced} by the
formation of a Wigner molecule within the dot, as shown by the study
of the transport dynamics of the lowest-lying excited states for
$N=6$.\\
\noindent The effects described above are due to the qualitative
rearrangements of the charge or spin patterns of the dot states,
occurring during the transition towards the Wigner molecule regime. As
a possible extension of this investigation, it would be interesting to
devise a method to investigate the internal spin structure of the
Wigner molecule, e.g. by analyzing the effects of spin-dependent
tunnel barriers. Effects due to Wigner molecules should have a
profound impact in coherent regimes and could lead to strong
signatures detected by analyzing e.g. the cotunnelling
regime. Finally, it would be interesting to consider the effects of
applied magnetic fields, which are known to strongly modify the
properties of Wigner molecules.\\
\noindent We expect that results similar to the ones shown in this
paper hold also for planar quantum dots and that they could be in
principle observed experimentally.  

\ack F. C. acknowledges financial support by CNR via Seed Project
PLASE001. U. D. G. acknowledges financial support by Fondazione Angelo
Della Riccia. We thank Dr. Achim Gelessus at the Computational
Laboratory for Analysis, Modeling and Visualization (CLAMV) of the
Jacobs University for the support on the cluster where numerical
calculations have been performed.

\appendix
\section{Tunnelling Hamiltonian}
\label{sec:a1}
In this appendix we derive $\hat{H}_{\mathrm t}$ for a quantum dot
connected to two external emitter or collector leads via tunnelling
barriers located around $z=z_{\alpha}$ ($\alpha=\mathrm{E,C}$), see
figure~\ref{fig:fig1}. Our starting point is the Bardeen tunnelling
Hamiltonian~\cite{Bardeen}
\begin{equation}
 \label{eq:tunbardeen}
\!\!\!\!\!\!\!\!\!\!\!\! \hat{H}_{\mathrm t}=\frac{i}{2m^{*}}\sum_{\alpha=\mathrm{E,C}}\int{\mathrm d}\mathbf{R}\ {\mathrm d}z\ \delta(z-z_{\alpha})\left[\hat{\Psi}^{\dagger}(\mathbf{R},z)\frac{\partial\hat{\Psi}(\mathbf{R},z)}{\partial z}-\mathrm{h.c.}\right]\, .
\end{equation}
Here, ${\mathbf R}=(x,y)$ and $z_{\alpha}$ is a point within the
tunnelling barrier between the lead $\alpha$ and the dot. Furthermore,
$\hat{\Psi}(\mathbf{R},z)$ is the system field operator expanded into
emitter, dot and collector contributions as
$\hat{\Psi}(\mathbf{R},z)=\hat{\Psi}_{\mathrm
  D}(\mathbf{R},z)+\sum_{\alpha=\mathrm{E,C}}\hat{\Psi}_{\alpha}(\mathbf{R},z)$ with
\begin{eqnarray}
\hat{\Psi}_{\alpha}(\mathbf{R},z)&=&\sum_{\xi_{\alpha}}\Phi_{\xi_{\alpha}}^{(\alpha)}(\mathbf{R},z)\hat{c}_{\alpha,\xi_{\alpha}}\\
\hat{\Psi}_{\mathrm{D}}(\mathbf{R},z)&=&\sum_{\eta}\Phi_{\eta}^{(\mathrm{D})}(\mathbf{R},z)\hat{d}_{\eta}\, .
\end{eqnarray}
Here, $\Phi_{\xi_{\alpha}}^{(\alpha)}(\mathbf{R},z)$ and
$\Phi_{\eta}^{(\mathrm{D})}(\mathbf{R},z)$ are a set of single
particle eigenfunctions for leads and dot, respectively. Note that
these sets may be complete but leads and dot states need not be
orthogonal. In the following, we assume that the longitudinal ($z$)
and transverse ($x,y$) motions are decoupled, which allows the
factorizations of the single particle WFs. Furthermore, we will
concentrate on a system with cylindrical symmetry. For the leads one
has
$\Phi_{\xi_{\alpha}}^{(\alpha)}(\mathbf{R},z)=\phi_{\nu_{\alpha}}(\mathbf{R})\chi_{k_{\alpha}}(z)$
where $k_{\alpha}$ is the momentum along the $z$ direction and
$\nu_{\alpha}$ is a set of quantum numbers describing the transverse
motion. For the quantum dot we choose
$\Phi_{\eta}^{(\mathrm{D})}(\mathbf{R},z)=\phi_{\eta}(\mathbf{R})\chi_{\mathrm{D}}(z)$,
where $\eta=\{n_{\eta}^{({\mathrm D})},l_{\eta}^{({\mathrm
    D})},s_{z\eta}^{({\mathrm D})}\}$ collectively denotes the FD
quantum numbers. Choosing the basis of FD states is not restrictive:
as shown in~\ref{sec:a3}, every orthonormal and complete basis for the
single particle states of the dot is equivalent. We assume a
sufficiently tight confinement in the $z$ direction so that
longitudinal motion of electrons in the dot is effectively frozen into
the lowest subband with WF $\chi_{\mathrm D}(z)$. The longitudinal dot
WF $\chi_{\mathrm D}(z)$ is evanescent within both tunnelling
barriers. For the leads, $\chi_{k_{\mathrm{E}}}(z)$
($\chi_{k_{\mathrm{C}}}(z)$) is evanescent under the tunnelling
barrier near the emitter (collector) and essentially zero near the
collector (emitter). Substituting the explicit expressions of the
field operators into~(\ref{eq:tunbardeen}) and taking into account the
form of the single particle WFs one obtains $\hat{H}_{\mathrm
  t}=\hat{H}_{\mathrm t}^{(0)}+\hat{H}_{\mathrm t}^{(1)}$ with
\begin{eqnarray}
\hat{H}_{\mathrm t}^{(0)}&=&\sum_{\alpha=\mathrm{E,C}}\sum_{\xi_{\alpha},\xi'_{\alpha}}\Delta_{\xi_{\alpha},\xi'_{\alpha}}^{(\alpha)}\hat{c}^{\dagger}_{\alpha,\xi_{\alpha}}\hat{c}_{\alpha,\xi'_{\alpha}}+\Delta^{(\mathrm{D})}\sum_{\eta}\hat{d}^{\dagger}_{\eta}\hat{d}_{\eta}\\
\hat{H}_{\mathrm t}^{(1)}&=&\sum_{\alpha=\mathrm{E,C}}\sum_{\xi_{\alpha},\eta}\tau_{\xi_{\alpha},\eta}^{(\alpha)}\hat{c}^{\dagger}_{\alpha,\xi_{\alpha}}\hat{d}_{\eta}+\mathrm{h.c.}\label{eq:tunhamold}\, ,
\end{eqnarray}
where
\begin{eqnarray}
  \Delta_{\xi_{\alpha},\xi'_{\alpha}}^{(\alpha)}&=&\frac{i}{2m^{*}}\delta_{\nu_{\alpha},\nu'_{\alpha}}\left[\chi^{*}_{k_{\alpha}}(z_{\alpha})\left.\frac{\partial\chi_{k'_{\alpha}}(z)}{\partial z}\right|_{z_{\alpha}}-\chi_{k'_{\alpha}}(z_{\alpha})\left.\frac{\partial\chi_{k'_{\alpha}}^{*}(z)}{\partial z}\right|_{z_{\alpha}}\right]\, ,\nonumber\\
  \Delta^{(\mathrm{D})}&=&\frac{i}{2m^{*}}\sum_{\alpha=\mathrm{E,C}}\left[\chi^{*}_{\mathrm{D}}(z_{\alpha})\left.\frac{\partial\chi_{\mathrm{D}}(z)}{\partial z}\right|_{z_{\alpha}}-\chi_{\mathrm{D}}(z_{\alpha})\left.\frac{\partial\chi_{\mathrm{D}}^{*}(z)}{\partial z}\right|_{z_{\alpha}}\right]\, ,\nonumber\\
  \tau_{\xi_{\alpha},\eta}^{(\alpha)}&=&\frac{i}{2m^{*}}\left[\chi^{*}_{k_{\alpha}}(z_{\alpha})\left.\frac{\partial\chi_{\mathrm{D}}(z)}{\partial z}\right|_{z_{\alpha}}-\chi_{\mathrm{D}}(z_{\alpha})\left.\frac{\partial\chi_{k_{\alpha}}^{*}(z)}{\partial z}\right|_{z_{\alpha}}\right]\cdot\nonumber\\
  &&\cdot\int{\mathrm d}\mathbf{R}\ \phi_{\nu_{\alpha}}^{*}(\mathbf{R})\phi_{\eta}(\mathbf{R})\label{eq:transamp}\, .
\end{eqnarray}
The first term in $\hat{H}_{\mathrm t}^{(0)}$ produces a weak,
one-body scattering within the leads, while the second term gives rise
to a small uniform shift of the dot energy levels. Both effects can be
safely neglected. The relevant term is $\hat{H}_{\mathrm t}^{(1)}$,
which produces scattering of electrons between the leads and the
dot. For a pillar quantum dot, assuming harmonic confinement of
electrons in the emitter and collector, with frequency
$\omega_{\alpha}$, the WFs $\phi_{\nu_{\alpha}}(\mathbf{R})$ are FD
states. We consider the case
$\omega_{\alpha}\approx\omega$. Equation~(\ref{eq:transamp}) yields
essentially
\begin{equation}
\label{eq:tau}
\tau_{\xi_{\alpha},\eta}=t_{k_{\alpha}}^{(\alpha)}\delta_{\nu_{\alpha},\eta}\approx t^{(\alpha)}\delta_{\nu_{\alpha},\eta}
\end{equation}
where in the last equality we have neglected the weak dependence of
the tunneling matrix element on $k_{\alpha}$. Finally, the tunnelling
Hamiltonian assumes the form
\begin{equation}
\label{eq:tunha1}
\hat{H}_{\mathrm t}=\sum_{\alpha}t^{(\alpha)}\sum_{\xi_{\alpha},\eta}\hat{c}^{\dagger}_{\alpha,\xi_{\alpha}}\hat{d}_{\eta}+\mathrm{h.c.}\, .
\end{equation}
\section{Tunnelling rates}
\label{sec:a2}
In the sequential regime, tunnelling rates between initial
$|\mathcal{I}\rangle$ and final $|\mathcal{F}\rangle$ states of the
{\em system} are obtained via the Fermi's golden
rule~\cite{Bruus}
\begin{equation}
\label{eq:therate}
\Gamma_{{\mathcal I}\to{\mathcal F}}=2\pi\left|\langle \mathcal{F}|\hat{H}_{\mathrm t}|\mathcal{I}\rangle\right|^{2}\delta(E_{\mathcal F}-E_{\mathcal I})\, ,
\end{equation}
where $E_{\mathcal I}$ ($E_{\mathcal F}$) is the total energy of the
system in the initial (final) state and
\begin{equation*}
|\mathcal{I}\rangle=|\mathcal{I}_{\mathrm{E}}\rangle\otimes|\mathcal{I}_{\mathrm {C}}\rangle\otimes|\mathcal{I}_{\mathrm{D}}\rangle\quad;\quad|\mathcal{F}\rangle=|\mathcal{F}_{\mathrm{E}}\rangle\otimes|\mathcal{F}_{\mathrm{C}}\rangle\otimes|\mathcal{F}_{\mathrm{D}}\rangle \, ,
\end{equation*}
where $|\mathcal{I}_{\alpha}\rangle$ ( $|\mathcal{F}_{\alpha}\rangle$)
is the initial (final) state for lead $\alpha$ while
$|\mathcal{I}_{\mathcal D}\rangle$ ($|\mathcal{F}_{\mathcal
  D}\rangle$) is the dot initial (final) state, with electron number
$N_{\mathcal{I}_{\mathrm D}}$ ($N_{\mathcal{F}_{\mathrm D}}$) and
energy $E_{{\mathcal I}_{\mathrm D}}$ ($E_{{\mathcal F}_{\mathrm
    D}}$). Note that dot states can be either ground or excited. In
order to have a nonvanishing contribution in the sequential regime,
$|N_{{\mathcal F}_{\mathrm D}}-N_{{\mathcal I}_{\mathrm D}}|=1$ must
hold. This implies that~(\ref{eq:therate}) is diagonal in the
barrier index $\alpha$ and that sequential tunnelling events through
the barriers are independent.  The tunnelling rate has the general
structure
\begin{equation*}
\Gamma_{\mathcal{I}\to\mathcal{F}}=\sum_{\alpha=\mathrm{E,C}}\sum_{p=\pm 1}\Gamma_{\mathcal{I}\to\mathcal{F}}^{(\alpha),p}\, , 
\end{equation*}
where the contribution with $p=+1$ ($p=-1$) represents tunnelling into
(out from) the dot via lead $\alpha$.\\
Since we are interested into the dot dynamics only, we perform a
thermal average over $|{\mathcal I}_{\alpha}\rangle$ and a summation
over $|{\mathcal F}_{\alpha}\rangle$ obtaining transition rates among
dot states only
\begin{equation}
\label{eq:tunhareda}
\Gamma_{\mathcal{I}_{\mathrm D}\to\mathcal{F}_{\mathrm D}}=\sum_{\alpha=\mathrm{E,C}}\sum_{p=\pm 1}\Gamma_{\mathcal{I}_{\mathrm D}\to\mathcal{F}_{\mathrm D}}^{(\alpha),p}\, . 
\end{equation}
The leads are assumed to be in equilibrium with respect to their
electrochemical potentials $\mu_{\alpha}=\mu_{0}+\delta\mu_{\alpha}$,
where $\delta\mu_{\alpha}$ is a shift due to the presence of an
applied bias voltage $V$. In the case of a pillar dot, see~(\ref{eq:tunha1}), one obtains
\begin{equation}
\Gamma_{\mathcal{I}_{\mathrm D}\to\mathcal{F}_{\mathrm
D}}^{(\alpha),p}=\Gamma^{(\alpha)}\left|\mathcal{O}_{p}\right|^{2}f_{p}(
\mu_{\mathrm D}-\mu_{\alpha})\, , 
\end{equation} 
where
$\mathcal{O}_{p}$:
\begin{equation}
\mathcal{O}_{1}=\sum_{\eta}\langle\mathcal{F}_{\mathrm D}|\hat{d}_{\eta}^{\dagger}|\mathcal{I}_{\mathrm D}\rangle\quad;\quad\mathcal{O}_{-1}=\sum_{\eta}\langle\mathcal{F}_{\mathrm D}|\hat{d}_{\eta}|\mathcal{I}_{\mathrm D}\rangle\, ,
\end{equation} 
and $f_{p}(E)=p f(E)+(1-p)/2$, with $f(E)=[1+\exp(\beta E)]^{-1}$ the
Fermi distribution at inverse temperature $\beta=1/k_{\mathrm B}T$
($k_{\mathrm B}$ the Boltzmann constant) and $\mu_{\mathrm
  D}=E_{{\mathcal F}_{\mathrm D}}-E_{{\mathcal I}_{\mathrm D}}$.
\section{Equivalence of different bases}
\label{sec:a3}
In this appendix we will show the equivalence of different single
particle bases of the dot for the definition of the tunneling
Hamiltonian. Consider two orthonormal and complete bases of single
particle dot states labeled by $\{\eta\}$ and $\{\mu\}$, with
corresponding orbitals $\phi_{\eta}(\mathbf{R})$,
$\bar{\phi}_{\mu}(\mathbf{R})$ and Fermi operators $d_{\eta}$,
$\bar{d}_{\mu}$ (in this section we omit overhats from the operators
to simplify the notation). They are connected by a unitary matrix
$M_{\eta\mu}$
\begin{equation}
\label{eq:optra}
\phi_{\eta}(\mathbf{R})=\sum_{\mu}M_{\eta\mu}\bar{\phi}_{\mu}(\mathbf{R})\ \ ;\ \ d_{\eta}=\sum_{\mu}M_{\eta\mu}^{\dagger}\bar{d}_{\mu}\, .
\end{equation}
The tunneling Hamiltonian, expressed in the basis $\{\mu\}$ reads
\begin{equation}
\label{eq:tunnew}
H_{\mathrm
  t}^{(1)}=\sum_{\alpha=\mathrm{E,C}}\sum_{\xi_{\alpha},\mu}\bar{\tau}_{\xi_{\alpha},\mu}^{(\alpha)}c^{\dagger}_{\alpha,\xi_{\alpha}}\bar{d}_{\mu}+\mathrm{h.c.}\,
,
\end{equation}
with 
\begin{eqnarray}
  \bar{\tau}_{\xi_{\alpha},\mu}^{(\alpha)}&=&\frac{i}{2m^{*}}\left[\chi^{*}_{k_{\alpha}}(z_{\alpha})\left.\frac{\partial\chi_{\mathrm{D}}(z)}{\partial z}\right|_{z_{\alpha}}-\chi_{\mathrm{D}}(z_{\alpha})\left.\frac{\partial\chi_{k_{\alpha}}^{*}(z)}{\partial z}\right|_{z_{\alpha}}\right]\cdot\nonumber\\
  &&\cdot\int{\mathrm d}\mathbf{R}\ \phi_{\nu_{\alpha}}^{*}(\mathbf{R})\bar{\phi}_{\mu}(\mathbf{R})\label{eq:transampnew}\, ,
\end{eqnarray}
see~(\ref{eq:transamp}). Inverting~(\ref{eq:optra}) for
$\bar{d}_{\mu}$ and plugging into~(\ref{eq:tunnew}) one obtains
\begin{equation}
\label{eq:tunnew2}
H_{\mathrm
  t}^{(1)}=\sum_{\alpha=\mathrm{E,C}}\sum_{\xi_{\alpha},\eta}\left[\sum_{\mu}M_{\eta\mu}\bar{\tau}_{\xi_{\alpha},\mu}^{(\alpha)}\right]c^{\dagger}_{\alpha,\xi_{\alpha}}d_{\eta}+\mathrm{h.c.}\,
.
\end{equation}
By virtue of~(\ref{eq:optra}), one can easily see that
\begin{equation}
\sum_{\mu}M_{\eta\mu}\bar{\tau}_{\xi_{\alpha},\mu}^{(\alpha)}\equiv\tau_{\xi_{\alpha},\eta}^{(\alpha)}
\end{equation}
which shows that~(\ref{eq:tunnew}) is identical
to~(\ref{eq:tunhamold}). This implies that all results are independent
of the choice of the single particle states for the dot.

\end{document}